\documentclass[a4paper,11pt,pdftex]{article}
\pdfoutput=1

\usepackage{slashed}
\usepackage{braket}
\usepackage{jheppub}
\usepackage{graphicx}
\usepackage{url}
\usepackage{amssymb,amsmath}
\usepackage{float}
\usepackage{wrapfig}
\usepackage{bm}
\usepackage{CJKutf8}
\usepackage{color}
\usepackage{comment}
\usepackage{xcolor}
\usepackage[normalem]{ulem}
\usepackage{hyperref}
\usepackage{mathrsfs}
\usepackage[version=4]{mhchem}
\usepackage{booktabs}
\usepackage[noabbrev]{cleveref}
\hypersetup{
colorlinks=true,
citecolor=blue,
citebordercolor=red,
linkcolor=blue,
urlcolor=blue
}

\catcode`\@=11
\catcode`\@=12

\newcommand{\p}{\partial}

\newcommand{\bp}{\begin{pmatrix}}
\newcommand{\ep}{\end{pmatrix}}

\newcommand{\del}{\partial}

\newcommand{\bs}[1]{\boldsymbol}

\newcommand{\be}{\begin{equation}}
\newcommand{\ee}{\end{equation}}

\newcommand{\ba}{\begin{array}} 
\newcommand{\ea}{\end{array}}

\graphicspath{{./figs/}}
\newbox{\ORCIDicon}

\makeatletter
\gdef\@fpheader{\phantom{prepared for submission to JHEP}}
\makeatother

\makeatletter
\@addtoreset{equation}{section}
\makeatother

\begin{document}
\begin{flushright} 
YGHP-24-07
\end{flushright} 
\title{
Domain-wall Skyrmion 
phase of QCD in magnetic field: 
Gauge field dynamics
}
\author[a,b]{Yuki Amari,}
\emailAdd{amari.yuki@keio.jp}

\author[c,a,d]{Minoru Eto,}
\emailAdd{meto@sci.kj.yamagata-u.ac.jp}

\author[b,a,d]{Muneto Nitta}
\emailAdd{nitta@phys-h.keio.ac.jp}

\affiliation[a]{Research and Education Center for Natural Sciences, Keio University, 4-1-1 Hiyoshi, Yokohama, Kanagawa 223-8521, Japan}

\affiliation[b]{Department of Physics, Keio University, 4-1-1 Hiyoshi, Yokohama, Kanagawa 223-8521, Japan}

\affiliation[c]{Department of Physics, Yamagata University, Kojirakawa-machi 1-4-12, Yamagata, Yamagata 990-8560, Japan}

\affiliation[d]{
International Institute for Sustainability with Knotted Chiral Meta Matter(SKCM$^2$), Hiroshima University, 1-3-2 Kagamiyama, Higashi-Hiroshima, Hiroshima 739-8511, Japan
}

\abstract{
The ground state of 
QCD in sufficiently strong magnetic field at finite baryon density is  
an inhomogeneous state consisting of 
an array of solitons, called 
the chiral soliton lattice (CSL). 
It is, however, replaced 
in a region with higher density and/or magnetic field 
by 
the so-called domain-wall Skyrmion(DWSk) phase 
where Skyrmions are created on top of the CSL. 
This was previously proposed within the Bogomol'nyi-Prasad-Sommerfield (BPS) approximation 
neglecting a gauge field dynamics and 
taking into account its effect by a flux quantization condition. 
In this paper, by taking into account dynamics of the gauge field, 
we show that the phase boundary 
between the CSL and DWSk phases 
beyond the BPS approximation 
is identical to the one obtained in the BPS approximation. 
We also find that domain-wall Skyrmions are electrically charged with the charge one as a result of the chiral anomaly. 
}

\maketitle 

\section{Introduction}

The phase diagram of
Quantum Chromodynamics (QCD) 
receives a quite extensive attention,
especially under extreme conditions like high baryon density, pronounced magnetic fields, and rapid rotation  \cite{Fukushima:2010bq}. 
In particular, 
strong magnetic fields have received quite intense attention because of their relevance in the interior of neutron stars
and heavy-ion collisions.
First principle calculation of QCD at finite density is difficult due to the sign problem.
On the other hand, at low energy QCD can be described model independently 
 by the chiral Lagrangian or the chiral perturbation theory (ChPT) 
\cite{Scherer:2012xha,Bogner:2009bt};
when chiral symmetry undergoes spontaneous breaking, there appear massless Nambu-Goldstone (NG) bosons or pions, which are dominant at low energy. The low-energy dynamics of QCD is governed by these light modes in terms of the aforementioned 
ChPT centered on the pionic degree of freedom. Importantly, this description is model independently  dictated by symmetries and only modulated by certain constants, including the pion's decay constant $f_{\pi}$ and quark masses $m_{\pi}$.   
Effects of external magnetic fields $B$
and finite chemical potential $\mu_{\textrm{B}}$ 
can be incorporated in the ChPT; 
It is accompanied by 
the Wess-Zumino-Witten (WZW) term containing  
an anomalous coupling of the neutral pion $\pi^0$ to the magnetic field via the chiral anomaly 
\cite{Son:2004tq,Son:2007ny}
in terms of 
the Goldstone-Wilczek current \cite{Goldstone:1981kk,Witten:1983tw},
determined to reproduce the so-called chiral separation effect 
\cite{Vilenkin:1980fu,
Son:2004tq,Metlitski:2005pr,
Fukushima:2010bq,Landsteiner:2016led} in terms of the neutral pion $\pi^0$.

 \begin{figure}[h]
    \begin{center}
    \includegraphics[width=10cm]{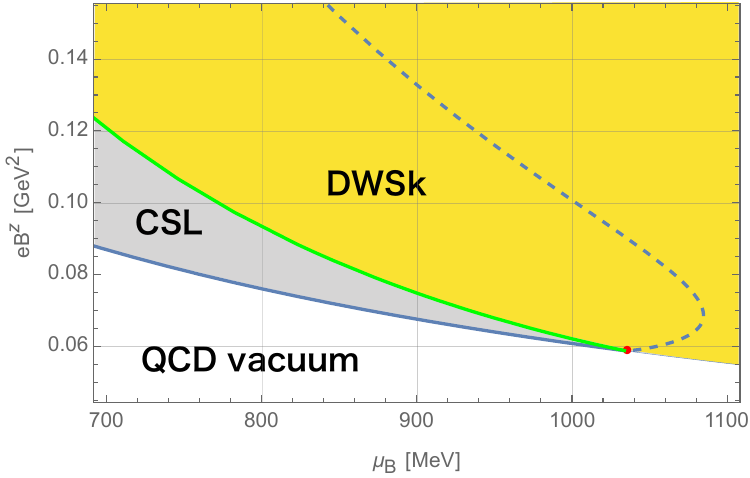}
    \caption{The phase diagram of QCD in the $\mu_{\rm B}$-$eB^z$
    plane. The blue solid and grey curves are the boundary with QCD vacuum above which CSL or DWSk phases are ground states, respectively. 
     The green curve is the phase boundary between the CSL and DWSk phases previously obtained in the BPS approximation, which we show in this paper is valid even taking into account gauge field dynamics.  
     The CSL state remains metastable between 
     the green and blue dotted curves, that becomes unstable beyond the blue dotted curve where the CPC occurs.
    The red point denotes a tricritical point 
    where the DWSk and CSL phases and QCD vacuum meet.}
    \label{fig:phase_diagram_AEN0}
    \end{center}
    \end{figure}

In this set up, 
it was found in Refs.~\cite{Son:2007ny,Eto:2012qd,Brauner:2016pko} that 
under a sufficiently strong magnetic field $B$ 
satisfying 
$B  
    \geq B_{\rm CSL} =
    {16\pi m_{\pi}/f_{\pi}^2}{e \mu_{\textrm{B}}}$, 
the ground state of QCD with two flavors  becomes inhomogeneous  
in the form of a chiral soliton lattice (CSL) 
consisting of a stack of 
domain walls or solitons 
carrying a baryon number (the blue solid and grey curves in Fig.~\ref{fig:phase_diagram_AEN0}). 
The quantum nucleation of such CSLs 
was studied in Refs.~\cite{Eto:2022lhu,Higaki:2022gnw},  
while the extension to include the $\eta'$ meson leads to quasicrystals \cite{Qiu:2023guy}.\footnote{ 
CSLs in 
QCD-like theory such as 
$SU(2)$ QCD and vector-like gauge theories were studied in 
Refs.~\cite{Brauner:2019rjg,Brauner:2019aid}, while those in supersymmetric QCD 
were studied in Ref.~\cite{Nitta:2024xcu}.
}
Although these results are based on 
zero temperature analyses, 
it was also shown that 
thermal fluctuations enhance their 
stability~\cite{Brauner:2017uiu,Brauner:2017mui,Brauner:2021sci,Brauner:2023ort}.
However,
such a CSL state was found to exhibit two kinds of instabilities, 
with replaced by another state.
One is 
a charged pion condensation (CPC) 
in a region of higher density and/or stronger magnetic field, 
asymptotically expressed 
at large $B$ as $B \geq B_{\rm CPC} 
    \sim {16 \pi^4 f_{\pi}^4}/{\mu_{\rm B}^2}$ \cite{Brauner:2016pko}  
above which the CSL becomes  unstable, 
and was proposed to go to 
 an Abrikosov's vortex lattice 
 or a baryon crystal \cite{Evans:2022hwr,Evans:2023hms}.
Another instability of the CSL 
relevant to our study occurring 
below the CPC instability 
corresponds to 
the appearance of 
a domain-wall Skyrmion (DWSk) phase in a region 
$B \geq B_{\rm DWSk}(< B_{\rm CPC})$ 
(denoted by the green curve in Fig.~\ref{fig:phase_diagram_AEN0})
in which Skyrmions are created on top of the solitons 
in the ground state \cite{Eto:2023lyo,Eto:2023wul}.\footnote{
In the case of QCD under rapid rotations, 
there appear CSLs composed on 
the $\eta$ or $\eta'$ meson ~\cite{Huang:2017pqe,Nishimura:2020odq,Chen:2021aiq,Eto:2021gyy,Eto:2023tuu,Eto:2023rzd}. In this case too, 
there is a DWSk phase \cite{Eto:2023tuu} 
similar to the case of magnetic fields.
}
These two instability curves meet 
at a single tricritical point
    $(\mu_{\rm c},B_{\rm c}) 
    = 
    \left({16\pi f_{\pi}^2}/{3m_{\pi}}, 
    {3m_{\pi}^2}/{e} \right) 
    \approx 
    \left(1.03 \;\; {\rm GeV}, 
    0.06 {\rm GeV}^2\sim 1.0\times 10^{19} {\rm G}\right)$ 
    (denoted by the red dot in Fig.~\ref{fig:phase_diagram_AEN0})
    on the critical curve 
    $ B= B_{\rm CSL}$ of the CSL phase.
 Skyrmions are topological solitons supported by the third homotopy group 
$\pi_3 [{\rm SU}(N_{\rm F}) ] \simeq  {\mathbb Z}$ 
in the chiral Lagrangian 
complemented by 
a four-derivative Skyrme term originally
proposed to describe baryons  \cite{Skyrme:1962vh} 
where $N_{\rm F}$ is a number of flavors, and in our context 
the interplay between Skyrmion crystals at zero magnetic field and the CSL 
was studied in Refs.~\cite{Kawaguchi:2018fpi,Chen:2021vou,Chen:2023jbq}.
The domain-wall Skyrmions are composite states of a domain wall and Skyrmions, initially introduced in the  field theoretical models in 
2+1 dimensions~\cite{Nitta:2012xq,Kobayashi:2013ju,Jennings:2013aea}\footnote{
In this dimensionality, a baby Skyrmion 
supported by $\pi_2(S^2) \simeq {\mathbb Z}$ in the bulk 
becomes a sine-Gordon soliton supported by 
$\pi_1(S^1) \simeq {\mathbb Z}$
in the domain-wall effective theory 
which is a sine-Gordon model.
In condensed matter physics such as chiral magnets,
domain-wall Skyrmions were theoretically investigated~\cite{PhysRevB.99.184412,KBRBSK,Ross:2022vsa,Amari:2023gqv,Amari:2023bmx,Gudnason:2024shv,
Leask:2024dlo,
PhysRevB.102.094402,Kim:2017lsi,Lee:2022rxi,Lee:2024lge,Amari:2024jxx} and were experimentally observed~\cite{Nagase:2020imn,Yang:2021}.
}
and in
3+1 dimensions~\cite{Nitta:2012wi,Nitta:2012rq,Gudnason:2014nba,Gudnason:2014hsa,Eto:2015uqa,Nitta:2022ahj}.
In QCD,  
Skyrmions 
in the bulk
are absorbed into a chiral soliton 
to become 
topological lumps (or baby Skyrmions) 
supported by $\pi_2(S^2)\simeq {\mathbb Z}$ in 
an O(3) sigma model 
or the ${\mathbb C}P^1$ model,
constructed by the moduli approximation \cite{Manton:1981mp,Eto:2006uw,Eto:2006pg}
as 
the effective worldvolume theory on a soliton \cite{Eto:2023lyo}. 
One of the important features is that one lump 
 in the soliton 
 corresponds to two Skyrmions in the bulk, 
 and thus they are bosons \cite{Amari:2024mip}.
Domain-wall Skyrmions in multiple chiral solitons 
(that is a CSL) 
are Skyrmion chains \cite{Eto:2023wul}, 
giving a more precise phase boundary between 
the CSL and DWSk phases.

In the previous works 
\cite{Eto:2023lyo,Eto:2023wul}, 
the so-called Bogomol'nyi-Prasad-Sommerfield (BPS) approximation was employed:
 lump solutions in the domain-wall effective theory are approximated by BPS lumps 
 with neglecting the gauge coupling, 
and subsequently taken into account the gauge field 
through a flux quantization condition.
On the other hand, in our previous paper
\cite{Amari:2024adu}, 
we constructed full gauged lump solutions 
in a gauged ${\mathbb C}P^1$ model 
and found that 
the flux quantization is 
only $15\%$ satisfied, 
and their energy (mass) are 
about $10\%$ less that that of the BPS lump.
We could expect a similar thing happens for domain-wall Skyrmions.

In this paper, we investigate the gauge field dynamics to construct gauged (anti-)lump solutions 
beyond the BPS approximation. 
First, we point out that anti-lumps in the 
domain-wall effective theory correspond to 
Skyrmions in the bulk, which can have negative energy due to the WZW term in the DWSk phase, 
while lumps in the wall correspond to anti-Skyrmions 
in the bulk 
which are always excited states.
We then find a critical gauge coupling above which 
a regular solution of gauged anti-lump solutions exist stably  
and below which solutions become singular. 
On contrary, gauged lumps (anti-Skyrmions) are always regular and stable. 
Our main results are twofold.
First, we find that gauged anti-lumps are singular 
in a realistic gauge coupling.
Nevertheless, their energy is finite 
coinciding with the BPS energy, 
and becomes negative in the DWSk phase. 
This implies that the phase boundary between 
the DWSk and CSL phases is unchanged from the BPS approximation. 
On the other hand, in strong gauge coupling, a region of regular solution expands and 
the DWSk phase expands in the phase diagram. 
Second, we find that gauged anti-lumps 
are electrically charged due to the chiral anomaly and their charges are one.

This paper is organized as follows.
In Sec.~\ref{sec:CSL} we review 
the chiral Lagrangian 
and the CSL phase. 
In Sec.~\ref{sec:DWSk}, 
the DWSk is constructed in terms of the domain-wall effective theory.
In Sec.~\ref{sec:phase-boundary}, 
the phase boundary between 
the DWSk and CSL phases 
and properties of domain-wall Skyrmions are investigated.
Section \ref{sec:summary} 
is devoted to a summary and discussion.

\section{The chiral soliton lattice in
QCD}\label{sec:CSL}

In this section, 
we give a review on chiral Lagrangian and CSL in order to fix our notations.

\subsection{
Chiral Lagrangian}
In this paper, we take the metric $\eta_{\mu\nu} = (+,-,-,-)$.
The chiral Lagrangian with 
$N_F=2$ flavors is given by
\be
{\cal L}_{\rm ChPT} = - \frac{1}{4} F_{\mu\nu}F^{\mu\nu} + \frac{f_\pi^2}{4}{\rm tr} D_\mu \Sigma^\dag D^\mu \Sigma + {\rm tr}\left(M\Sigma + \text{h.c.}\right)\,,
\label{eq:Lag_ChPT}
\ee
where pions are parameterized by
\be
\Sigma = \exp\left(\frac{i\tau^a\varphi^a}{f_\pi}\right) = \frac{1}{f_\pi}\left(\sigma + i \tau^a\pi^a\right)\,,\quad\left(\sigma^2 + \pi^a\pi^a = f_\pi^2\right)\,,
\label{eq:Sigma_pion}
\ee
the covariant derivative is given by
\be
D_\mu \Sigma = \p_\mu \Sigma + i e A_\mu\left[Q,\Sigma\right]\,,\qquad
Q = \frac{\tau^3}{2} + \frac{1}{6}{\bf 1}_2\,,
\ee
and $M$ is the pion mass matrix explicitly given below.
The charged pions are identified by 
$\pi^\pm \equiv \pi^1 \mp i \pi^2$, and the covariant derivative on them is $D_\mu \pi^\pm = \left(\p_\mu \pm i e A_\mu\right)\pi^\pm$.
The $U(1)_{\rm EM}$ electromagnetic gauge transformation is given by 
\be
\Sigma \to e^{iQ\alpha(x)} \Sigma e^{-iQ\alpha(x)}~
, \quad
\pi^\pm \to e^{\pm i\alpha(x)}\pi^\pm\,,
\quad
A_\mu \to A_\mu - e^{-1} \p_\mu \alpha(x)\,.
\label{eq:EM_transf}
\ee

In addition to ${\cal L}_{\rm ChPT}$, 
the WZW term is needed to reproduce the chiral anomaly. 
For the two flavors, $N_F = 2$, it is given by
\be
{\cal L}_{\rm WZW} = 
- A_\mu^{\rm B} j_{\rm B}^\mu 
+ \frac{\epsilon^{\mu\nu\alpha\beta}}{16\pi^2}
\left(
\frac{1}{3}eA_\mu {\rm tr}\left(L_\nu L_\alpha L_\beta\right) - \frac{ie^2}{4}F_{\mu\nu}A_\alpha {\rm tr}\left[\tau^3\left(L_\beta + R_\beta\right)\right]
\right)\,,
\ee
where 
$A_\mu^{\rm B}$ 
is a baryon gauge field, 
and 
$j_{\rm B}^\mu$  is the Goldstone-Wilczek  baryon number current, given by
\be
j_{\rm B}^\mu = - \frac{1}{24\pi^2} \epsilon^{\mu\nu\alpha\beta}
\left\{
{\rm tr}\left( L_\nu L_\alpha L_\beta \right)
- \frac{3 i e}{2} \p_\nu \left[
A_\alpha {\rm tr}\left[\tau^3\left(L_\beta + R_\beta\right)\right]
\right]
\right\}\,,
\label{eq:jB}
\ee
with
\be
L_\mu = \Sigma\p_\mu \Sigma^\dag\,,\quad
R_\mu = \p_\mu \Sigma^\dag\Sigma\,.
\ee
Our notation of the totally anti-symmetric tensor is $\epsilon^{0123} = - \epsilon_{0123} = 1$.
After short algebras, the $N_F=2$ WZW term can be expressed as \cite{Son:2007ny}
\be
{\cal L}_{\rm WZW} = -  A_\mu^{\rm B} j^\mu_{\rm B} - \frac{e}{2}A_\mu j^\mu_{\rm B}\,.
\label{eq:Lag_WZW}
\ee
Note that the baryon number in our notation is given by
\be
N_{\rm B} = \int d^3x\, (- j_{\rm B}^0)\,.
\ee
Thus, the $N_{\rm F}=2$ chiral Lagrangian we will work on in this paper can be summarized as
\be
{\cal L}_{\rm ChPT+WZW} = \frac{f_\pi^2}{4}{\rm tr} D_\mu \Sigma^\dag D^\mu \Sigma + {\rm tr}\left(M\Sigma + \text{h.c.}\right)
- \left(A_\mu^{\rm B} + \frac{e}{2}A_\mu\right)j^\mu_{\rm B}\,.
\label{eq:Lag_ChPT+WZW}
\ee
The energy functional corresponding to 
this Lagrangian is for static configurations:
\be
{\cal M} = \frac{1}{4} F_{ij}^2 + \frac{f_\pi^2}{4}{\rm tr} D_i \Sigma^\dag D_i \Sigma - {\rm tr}\left(M\Sigma + \text{h.c.}\right)
+ \left(A_\mu^{\rm B} + \frac{e}{2}A_\mu\right)j^\mu_{\rm B}.
\label{eq:energy_functional}
\ee

We turn on a uniform background magnetic field $B_i = \frac{1}{2}\epsilon_{ijk}F^{jk}$.
For ease of notation, we set the magnetic field being parallel to the $z=x^3$ axis as
\be
(F^{23},F^{31},F^{12}) = (0,0,-B^z)
\quad \Leftrightarrow \quad
(A_0,A_1,A_2,A_3) = \left(0,\frac{B^z}{2}x^2,-\frac{B^z}{2}x^1,0\right)\,.
\label{eq:BG_A}
\ee
In addition, we introduce a non-zero baryon chemical potential through the temporal component of the baryonic $U(1)$ gauge field with the chemical potential $\mu_{\rm B}$ as
\be
A^{\rm B}_{\mu} = (\mu_{\rm B},0,0,0)\,,\quad \mu_{\rm B} \ge 0\,.
\label{eq:BG_gauge}
\ee

\subsection{The chiral soliton lattice}

The CSL appears as the ground state in the presence of the sufficiently large   background magnetic field and the finite baryon chemical potential. 
In order to show CSL solutions, it is sufficient to
set $\varphi^1 = \varphi^2 = 0$, in which  $\Sigma$ reduces to 
\be
\Sigma = \exp\left(\frac{i}{f_\pi}\tau^3\varphi^3\right)\,,\quad    L_\mu = R_\mu = - \frac{i}{f_\pi}\tau^3\p_\mu\varphi^3\,.
\ee
Then, the Goldstone-Wilczek current can be expressed as
\begin{eqnarray}
    j_{\rm B}^\mu = \frac{e\epsilon^{\mu\nu\alpha\beta}}{4f_\pi\pi^2}\p_\nu \left(A_\alpha\p_\beta\varphi^3\right)
    = \frac{e\epsilon^{\mu\nu\alpha\beta}}{4f_\pi\pi^2} \left(\p_\nu A_\alpha\right)\p_\beta\varphi^3\,.
\end{eqnarray}
Assuming that $\varphi^3$ depends on the $x^3$ coordinate only, we have
\begin{eqnarray}
j_{\rm B}^{0} 
= \frac{e\epsilon^{0ijk}}{4f_\pi\pi^2} \left(\p_i A_j\right)\p_k\varphi^3
= - \frac{e}{4\pi^2 f_\pi} B^z \p_3 \varphi^3\,,
\end{eqnarray}
where we have used Eq.~(\ref{eq:BG_gauge}) and our notation is $\epsilon^{123} = - \epsilon_{123} = 1$.
Then the chiral Lagrangian reduces to
\be
{\cal L}_{{\rm ChPT}+{\rm WZW}} = \frac{f_\pi^2}{2}\p_\mu\chi^3\p^\mu\chi^3 + f_\pi^2 m_\pi^2 \cos \chi^3 + \frac{eB^z\mu_{\rm B}}{4\pi^2} \frac{\p\chi^3}{\p x^3}\,,
\label{eq:EOM_reduced}
\ee
where we have expressed the pion mass
\be
M = \frac{f_\pi^2m_\pi^2}{4} {\bf 1}_2 \,,
\ee
and $\chi^3$ is a dimensionless field defined by
\be
\varphi^3 = f_\pi \chi^3\,,
\ee
The Lagrangian in Eq.~(\ref{eq:EOM_reduced})
is known as a chiral sine-Gordon model.

The equation of motion (EOM) for the above reduced Lagrangian (\ref{eq:EOM_reduced}) reads 
\be
\p_\mu\p^\mu \chi^3 + m_\pi^2 \sin \chi^3 = 0\,.
\label{eq:EOM_pi0}
\ee
This is simply sine-Gordon equation. An analytic solution  for multiple solitons, called the CSL, is 
\be
\chi^3_{\rm CSL}(\zeta) = 2\, {\rm am}\left(\frac{m_\pi \zeta}{\kappa},\kappa\right) + \pi\,,
\label{eq:CSL}
\ee
where $\zeta$ is defined by $\zeta = \vec k \cdot \vec x$ with $\vec k$ being a unit vector that specifies the normal direction to the solitons.
Here, $\kappa$ is a real number within $0 \le \kappa \le 1$, 
called the elliptic modulus. 
The elliptic modulus is related to
a period $\ell$ of the CSL as
\be
\ell = \frac{2 \kappa K(\kappa)}{m_\pi}\,,
\ee
where $K$ is the complete elliptic integral of the first kind. 
A single sine-Gordon soliton corresponds to the limit $\kappa \to 1$ in which the period becomes infinite $\ell \to \infty$.

\begin{figure}[h]
\begin{center}
\includegraphics[width=10cm]{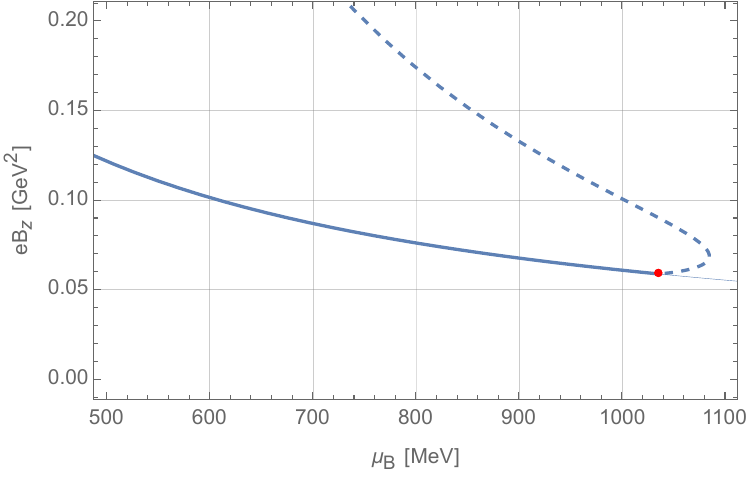}
\caption{The homogeneous QCD vacuum is below the solid curve corresponding to the single CS limit ($\kappa \to 1$) with $\mu_{\rm B} = \mu_{\rm B}^{(1)}$ given in Eq.~(\ref{eq:PB_SS}). The dashed curve corresponds to $B^z = B^{z(2)}$ given in Eq.~(\ref{eq:B_BY}). The CSL is (meta)stable between the solid and dashed curves. The red point is located at $\left(\dfrac{16\pi f_\pi^2}{3m_\pi},3m_\pi^2\right)$. We use $m_\pi = 140$ MeV and $f_\pi = 93$ MeV for an illustration.}
\label{fig:SSBY}
\end{center}
\end{figure}

The tension of the single soliton (one period of CSL) is given by
\begin{eqnarray}
\sigma_{\rm CSL} 
= 4m_{\pi}f_{\pi}^2\left[
    \frac{2E(\kappa)}{\kappa} + \left(\kappa-\frac{1}{\kappa} \right)K(\kappa)
    \right] - \frac{ek^3B^z\mu_{\rm B}}{2\pi}\,.
    \label{eq:M_CSL0}
\end{eqnarray}
The second term is minimized by $k^3 = 1$ and the solitons are perpendicular to the $x^3$ axis (the direction of the magnetic field).
In what follows, we fix $\vec k = (0,0,1)$.
To determine optimized $\kappa$ for given $B^z$ and $\mu_{\rm B}$, we minimize the mean tension $\sigma_{\rm CSL}/\ell$. The condition is
\begin{eqnarray}
    \frac{d}{d\kappa}\frac{\sigma_{\rm CSL}}{\ell}
    = \frac{m_\pi E(\kappa) \left(16\pi f_\pi^2 m E(\kappa) - e B^z \kappa \mu_{\rm B}\right)}{4\kappa^3\left(\kappa^2-1\right) \pi K(\kappa)^2}
     = 0\,.
\end{eqnarray}
This is satisfied by
\be
\frac{E(\kappa)}{\kappa} = \frac{eB^z\mu_{\rm B}}{16\pi f_\pi^2 m_\pi}\,.
\label{eq:minimizing_cond}
\ee
Thus, the minimum tension of a single soliton is
\be
\sigma_{\rm CSL}(\kappa) 
    = 4m_{\pi}f_{\pi}^2 \left(\kappa-\frac{1}{\kappa} \right)K(\kappa)\,.
    \label{eq:M_CSL}
\ee
This is always negative so that CSL is energetically more favourable than the homogeneous QCD vacuum for any $\kappa$ in $0 \le \kappa \le 1$.
Note that the elliptic integral satisfies $E(\kappa)/\kappa \ge 1$. Combining this with Eq.~(\ref{eq:minimizing_cond}), we find the condition that CSL is the ground state 
\be
\mu_{\rm B} \ge \frac{16\pi m_\pi f_\pi^2}{eB^z}\equiv \mu_
{\rm B}^{(1)}\,.
\label{eq:PB_SS}
\ee
The bound is saturated for the single soliton at $\kappa \to 1$, 
denoted the blue solid curves in Figs.~\ref{fig:SSBY} 
and \ref{fig:phase_diagram_AEN0}.

Note that the CSL solution given in Eq.~(\ref{eq:CSL}) corresponds to a contractible loop in the $SU(2)$ manifold, and thus it is  unstable against local fluctuations in the absence of the background magnetic field. The local stability problem (condensation of the charged pions) was found to be equal to the Landau level problem, and the bound for the background magnetic field is given by \cite{Son:2007ny,Brauner:2016pko}
\be
eB^z \ge \frac{m_\pi^2}{\kappa^2}
\left(2-\kappa^2 + 2\sqrt{1-\kappa^2+\kappa^4} \right)\equiv eB^{z(2)}\,.
\label{eq:B_BY}
\ee
The threshold value of $\mu_{\rm B}$ corresponding to $B^{z(2)}$ is implicitly given by $\kappa$ through Eq.~(\ref{eq:minimizing_cond}), 
denoted by 
the blue dotted curve in Figs.~\ref{fig:SSBY} and \ref{fig:phase_diagram_AEN0}, the phase diagram in the $\mu_{\rm B}$-$eB^z$ plane.

\section{The effective theory of non-Abelian chiral solitons}
\label{sec:DWSk}
In the previous section, we have reviewed the CSL, an infinite array of the flat solitons perpendicular to the $x^3$ axis. Both global and local stabilities were clarified, but there is a loophole corresponding to possibility that a topologically nontrivial excitation arises on the flat solitons.
Usually, a topological excitation costs some energy compared to a homogeneous vacuum (corresponding to the flat solitons in our case), so one might think such topologically nontrivial states do not change stability of CSL. However, it is not the case, because the WZW term in the presence of nonzero background magnetic field 
contributes the energy negatively.

\subsection{Non-Abelian chiral soliton lattice}

In order to include topologically nontrivial excitation on CSL, we first construct a low energy effective theory on CSL by using a standard moduli approximation method.

If we turn off the electromagnetic charge of the pions, the vector-like $SU(2)$ acting on
\be
\Sigma \to g \Sigma g^\dag\,,\quad g \in SU(2)\,
\ee
is genuine symmetry of the model.
Then we can produce new CSL solutions by taking
\be
\Sigma_0 = \exp\left(i\tau^3 \chi_{\rm CSL}^3\right)\,.
\ee
The solution can be parameterized by two component complex vector $\phi$ as follows
\be
\Sigma = g  \Sigma_0 g^\dag
= \exp\left(i g\tau^3g^\dag \chi_{\rm CSL}^3\right)
= \exp\left(2i\phi\phi^\dag \chi_{\rm CSL}^3\right) u^{-1}\,,
\label{eq:Sigma_g}
\ee
where $u$ is defined by
$u = e^{i\chi_{\rm CSL}^3}$,
and $\phi^T = (\phi_1,\phi_2)$ is given by $g$ as
\begin{align}
    g \tau_3 g^{\dag} = 2 \phi \phi^{\dag} - \bm{1}_2\,.
\end{align}
Note that $\phi$ satisfies $\phi^{\dag}\phi = 1$, and its phase is ambiguous. Thus the new solutions are one to one correspondence to $S^3/S^1 \simeq \mathbb{C}P^1$.
By using the following identity
\begin{eqnarray}
\exp\left(2i\phi\phi^\dag \chi^3_{\rm CSL}\right)
= {\bf 1} + (u^2 - 1)\phi\phi^\dag\,,
\end{eqnarray}
$\Sigma$ can be reexpressed as
\be
\Sigma 
= \left[{\bf 1} + (u^2 - 1)\phi\phi^\dag\right]u^{-1}\,.
\label{eq:Sigma}
\ee
Of course, $\phi$ is the genuine moduli parameter only when $B^z = 0$. It is only approximately moduli parameter for $B^z \neq 0$.

Let us promote the moduli parameter $\phi$ of the CSL to an effective field on the worldvolume of CSL as
$\phi = \phi(x^\alpha)$ ($\alpha=0,1,2$), and derive a low energy effective action. 
The gauge field is assumed to depend on only $x^\alpha$ as
\be
A_{0,3} = a_{0,3}(x^\alpha)\,,\quad
A_{1,2} = {\cal A}_{1,2}(x^1,x^2) + a_{1,2}(x^\alpha)\,,
\ee
where ${\cal A}_{1,2}$ is the background configuration
\be
{\cal A}_a = \frac{B^z}{2}\epsilon_{ab}x^b\,,\quad (a,b=1,2)\,.
\label{eq:A_bkg}
\ee
In the following, we shall take the axial gauge 
\be
A_3 = a_3 = 0\,.
\ee

\subsection{The effective theory 
of the soliton worldvolume}

Here, we construct the effective 
worldvolume theory on the solitons  
by the moduli approximation,  
sometimes called the Manton approximation 
\cite{Manton:1981mp,Eto:2006pg,Eto:2006uw}. 
In this method, we promote the moduli parameters of the solitons to fields 
on the worldvolume $(x^0,x^1,x^2)$, 
and integrate out the codimension $x^3$.

\subsubsection{The chiral Lagrangian part}

The covariant derivatives of $\Sigma$ read
\begin{eqnarray}
D_3 \Sigma &=&
g e^{i\tau^3\chi_{\rm CSL}^3}i\tau^3\p_3\chi_{\rm CSL}^3 g^\dag
\,,\\
D_\alpha \Sigma &=&
(u-u^{-1})\left((D^\alpha \phi)\phi^\dag + \phi D^\alpha\phi^\dag\right)\,,
\end{eqnarray}
with
\be
D_\alpha \phi = \p_\alpha \phi + i \frac{e}{2}A_\alpha \tau^3 \phi\,,\quad
D_\alpha \phi^\dag = \p_\alpha \phi^\dag - i \frac{e}{2} \phi^\dag \tau^3 A_\alpha\,,
\ee
and the electromagnetic 
$U(1)_{\rm EM}$ gauge transformation is
\be
\phi \to e^{i\frac{\tau^3}{2}\alpha(x)}\phi\,.
\label{eq:EM_transf2}
\ee
It is straightforward to obtain the following
\begin{eqnarray}
{\rm tr}D_3\Sigma^\dag D^3\Sigma 
 &=& 2 \p_3\chi_{\rm CSL}^3 \p^3\chi_{\rm CSL}^3\,,\\
 {\rm tr}D_\alpha\Sigma^\dag \p^\alpha\Sigma &=& - 2(u-u^{-1})^2\left[D_\alpha \phi^\dag D^\alpha \phi + (\phi^\dag D_\alpha\phi) (\phi^\dag D^\alpha \phi)\right]\,.
\end{eqnarray}
Plugging these into the chiral Lagrangian (\ref{eq:Lag_ChPT}), and integrating it over $x^3$ for  one period $\ell$, we find
\begin{eqnarray}
{\cal L}_{\rm ChPT}^{\rm eff} 
&=& \int^\ell_0 dx^3\, \left\{\frac{f_\pi^2}{4}{\rm tr} D_\mu \Sigma^\dag D^\mu \Sigma + {\rm tr}\left(M\Sigma + \text{h.c.}\right)\right\} \nonumber\\
&=& - 4m_{\pi}f_{\pi}^2\left[
    \frac{2E(\kappa)}{\kappa} + \left(\kappa-\frac{1}{\kappa} \right)K(\kappa)
    \right] \nonumber\\
&&+~ {\cal C}(\kappa)   
\left[D_\alpha \phi^\dag D^\alpha \phi + (\phi^\dag D_\alpha\phi) (\phi^\dag D^\alpha \phi)\right]\,,
\label{eq:L_eff_ChPT}
\end{eqnarray}
where we have defined the K\"{a}hler class
\begin{eqnarray}
    {\cal C}(\kappa) \equiv -\frac{f_\pi^2}{2} \int^\ell_0 dx^3\, (u-u^{-1})^2 
    =\frac{16 f_\pi^2}{3m_\pi} \beta(\kappa)\,,
\end{eqnarray}
with
\be
\beta(\kappa) \equiv 
\frac{(2-\kappa^2)E(\kappa)-2(1-\kappa^2)K(\kappa)}{\kappa^3}\,.
\label{eq:beta}
\ee
Note that the first term in the final expression of Eq.~(\ref{eq:L_eff_ChPT}) is constant and it can be rewritten by the tension of background CSL given in Eq.~(\ref{eq:M_CSL}) as
$4m_{\pi}f_{\pi}^2\left[
    \frac{2E(\kappa)}{\kappa} + \left(\kappa-\frac{1}{\kappa} \right)K(\kappa)
    \right] = \sigma_{\rm CSL}(\kappa) + \frac{eB^z\mu_{\rm B}}{2\pi}$.
Thus, the chiral Lagrangian gives
\begin{eqnarray}
{\cal L}_{\rm ChPT}^{\rm eff} 
= - \sigma_{\rm CSL}(\kappa) - \frac{eB^z\mu_{\rm B}}{2\pi}
+ {\cal C}(\kappa)   
\left[D_\alpha \phi^\dag D^\alpha \phi + (\phi^\dag D_\alpha\phi) (\phi^\dag D^\alpha \phi)\right]\,.
\label{eq:L_eff_ChPT2}
\end{eqnarray}
This is a $U(1)$ gauged ${\mathbb C}P^1$ 
model.

\subsubsection{The WZW term}

Let us next compute a contribution from the WZW term in Eq.~(\ref{eq:Lag_WZW}).
We first calculate the first term of Eq.~(\ref{eq:Lag_WZW}).
Since $A_0^{\rm B} = \mu_{\rm B}$ is the only nonzero component of $A_\mu^{\rm B}$, we only need $j_{\rm B}^0$.
Let us decompose $j^0_{\rm B}$ given in Eq.~(\ref{eq:jB}) into two parts as
\be
j_{\rm B}^0 = - {\cal B} + \tilde{\cal B}\,,
\label{eq:JB0}
\ee
with
\begin{eqnarray}
{\cal B} 
&\equiv& \frac{1}{24\pi^2} \epsilon^{ijk}
{\rm tr}\left( L_i L_j L_k \right)\,,
\label{eq:B_density}
\\
\tilde {\cal B} 
&\equiv& \frac{ie}{16\pi^2}\epsilon^{ijk}\p_i \left(
A_j {\rm tr}\left[\tau^3\left(L_k + R_k\right)\right]
\right)\,,
\label{eq:tB_density}
\end{eqnarray}
where we have used $\epsilon^{0ijk} = \epsilon^{ijk}$.
Here $\tilde {\cal B}$ 
is the baryon number density for chiral solitons, 
while ${\cal B}$ is the baryon number density for Skyrmions whose spatial integration corresponds to the Skyrmion baryon number
\be
N_{\rm B}^{\rm Sk} = 
\int d^3x\, {\cal B}\,.
\ee 

We then express $j^0_{\rm B}$ in terms of the moduli field $\phi$. To this end, let us introduce a two by two projection matrix
\be
P = \phi\phi^\dag\,,\quad
P^\dag = P\,,\quad
P^2 = P\,.
\ee
Then, $\Sigma$ in Eq.~(\ref{eq:Sigma}) can be expressed as
\be
\Sigma = {\bf 1}u^{-1} + (u - u^{-1})P \,.
\label{eq:Sigma_u}
\ee
Using $u = e^{i\chi_{\rm CSL}^3}$,
$\p_i u = i\p_i\chi_{\rm CSL}^3 u$, and
$\p_i u^{-1} = - i \p_i\chi_{\rm CSL}^3 u^{-1}$,
we can show
\begin{eqnarray}
L_i = i\p_i\chi_{\rm CSL}^3({\bf 1}-2P) - (1-u^{-2})\p_i P - (u-u^{-1})^2 P \p_i P\,,
\end{eqnarray}
for $i=1,2,3$.
Similarly, we have
\be
R_i = i\p_i\chi_{\rm CSL}^3({\bf 1}-2P) - (1-u^{-2})\p_i P - (u-u^{-1})^2 \p_i P P \,.
\ee
Plugging these into Eq.~(\ref{eq:B_density}), we obtain
\begin{eqnarray}
    {\cal B} 
    = \frac{1}{8\pi^2}{\rm tr}[L_1,L_2]L_3 
    = \frac{(u-u^{-1})^2\p_3\chi_{\rm CSL}^3}{2\pi} \times q\,,
\end{eqnarray}
where we have defined
\be
q \equiv 
- \frac{i}{2\pi} {\rm tr}\left[\p_1 P,\p_2 P\right]P
= - \frac{i}{2\pi}\epsilon^{ab}\p_a\phi^\dag\p_b\phi\,,
\label{eq:lump_charge_q}
\ee
which is the topological lump number density associated with $\pi_2(\mathbb{C}P^1)\simeq {\mathbb Z}$,
such that
\be
N_{\rm lump} \equiv \int d^2x\, q \in \mathbb{Z}\,.
\ee
For the CSL solution,
we have
\be
\int_0^{\ell} dx^3\, (u-u^{-1})^2\p_3\chi^3_{\rm CSL} = -4\pi\,.
\ee
Therefore, we have
\be
\int^\ell_0 dx^3\, {\cal B}
= - 2 q\,.
\label{eq:int_B}
\ee
By integrating this over $x^1$ and $x^2$, we find a relation between the baryon and lump numbers as\footnote{
Note the presence of the minus sign in this relation, which was missed in the previous works. 
Thus, anti-lumps correspond to baryons while lumps do to anti-baryons.
}
\be
N_{\rm B}^{\rm Sk} = 
\int d^3x\, {\cal B} = - 2 N_{\rm lump} \in 2 \mathbb{Z}\,.
\label{eq:Nb_Nl}
\ee

Next we compute $\tilde{\cal B}$. To this end, we introduce $M_k$ by
\be
M_k \equiv {\rm tr}\left[\tau^3\left(L_k + R_k\right)\right] = \left\{
\begin{array}{ccl}
- (u^2 - u^{-2}) {\rm tr}\left(\tau^3\p_k P\right) & & k=1,2\\
-4 i \p_3 \chi^3\, {\rm tr}(\tau^3P) && k = 3
\end{array}
\right.
\,,
\ee
then $\tilde{\cal B}$ in Eq.~(\ref{eq:tB_density})  can be expressed as
\be
\tilde{\cal B} = \frac{ie}{16\pi^2}
\left[
F_{12} 
M_3 - (\p_2M_3-\p_3M_2)A_1 - (\p_3 M_1 - \p_1 M_3)A_2\right]\,.
\label{eq:tilde_B}
\ee
The right hand side can explicitly be written in terms of $\phi$ as
\begin{eqnarray}
    \p_2M_3-\p_3M_2 
    &=& -4i   {\rm tr}\left[
    \tau^3
    \p_2 P
    \right] \p_3\chi^3_{\rm CSL}
    +  {\rm tr}\left[\tau^3\p_2 P \right] \p_3\left(u^2 - u^{-2}
    \right)\,,\\
    \p_3 M_1 - \p_1 M_3 
    &=& 
    4i   {\rm tr}\left[
    \tau^3
    \p_1 P
    \right] \p_3\chi^3_{\rm CSL}
    -  {\rm tr}\left[\tau^3\p_1 P \right] \p_3\left(u^2 - u^{-2}
    \right)\,.
\end{eqnarray}
Integrating these over a one period $\ell$ of CSL, we have
\begin{eqnarray}
\int_0^\ell dx^3\, F_{12} M_3 &=& -8\pi i F_{12} \, {\rm tr}\left(\tau^3 P\right)\,,\\
\int_0^\ell dx^3\, \left(\p_2M_3-\p_3M_2\right)A_1 
&=& -8\pi i \, {\rm tr}\left[\tau^3\p_2 P\right]A_1\,,\\
\int_0^\ell dx^3\, \left(\p_3 M_1 - \p_1 M_3\right)A_2 &=& 8\pi i \, {\rm tr}\left[\tau^3\p_1 P\right]A_2\,,
\end{eqnarray}
where we have used 
$\int^\ell_0dx^3\, \p_3\chi^3 = 2\pi$, 
$\int^\ell_0dx^3\, \p_3 ( u^2 - u^{-2}) = 0$,
$A_3 = 0$, and the fact that $A_{1,2}$ does not depend on $x^3$.
We thus find
\begin{eqnarray}
\int^\ell_0 dx^3\,  \tilde{\cal B} 
= \frac{e}{2\pi} 
\left(
F_{12} - A_1\p_2 + A_2\p_1
\right)
{\rm tr}\left[\tau^3 P\right]
= \frac{e}{2\pi} \epsilon^{ab}\p_a\left(A_b{\rm tr}\left[\tau^3 P\right]\right)\,.
\label{eq:int_tilde_B}
\end{eqnarray}
Combining Eqs.~(\ref{eq:int_B}) and (\ref{eq:int_tilde_B}), a contribution from the first term of the WZW term to the effective Lagrangian reads
\begin{eqnarray}
\int^\ell_0 dx^3\,\left(-A_\mu^{\rm B} j_{\rm B}^\mu\right) &=& - \mu_{\rm B} \int^\ell_0 dx^3\, \left(- {\cal B} + \tilde {\cal B}\right) \nonumber\\  
&=& - \mu_{\rm B} \left\{2 q 
+ \frac{e}{2\pi} 
\epsilon^{ab}\p_a\left(A_b{\rm tr}\left[\tau^3 P\right]\right)
\right\}\,.
\label{eq:L_eff_WZW}
\end{eqnarray}

There is another contribution from the second term of Eq.~(\ref{eq:Lag_WZW}). The contribution from $A_0$ is the same as that of $A_0^{\rm B}$, so we immediately find
\begin{eqnarray}
\int^\ell_0 dx^3\,\left(-\frac{e}{2}A_0 j_{\rm B}^0\right)
= - \frac{e}{2}A_0 \left\{2q 
+ \frac{e}{2\pi} 
\epsilon^{ab}\p_a\left(A_b{\rm tr}\left[\tau^3 P\right]\right)
\right\}.
\label{eq:L_eff_WZW2}
\end{eqnarray}
There is no contribution of $A_3$ since we set $A_3 = 0$.
For a contribution of $A_a$ $(a=1,2)$, one needs to compute $j_{\rm B}^a$ but it turns out to be zero as
\be
j_{\rm B}^{a} = - \frac{1}{24\pi^2} \epsilon^{a\nu\alpha\beta}
\left\{
{\rm tr}\left( L_\nu L_\alpha L_\beta \right)
- \frac{3 i e}{2} \p_\nu \left[
A_\alpha {\rm tr}\left[\tau^3\left(L_\beta + R_\beta\right)\right]
\right]
\right\} = 0\,,
\ee
because of $A_3 = 0$ and $\p_3 = 0$.
Eq.~(\ref{eq:L_eff_WZW2})
implies that 
an electric charge is induced 
around a lump, 
mostly proportional to its lump charge 
density. 
This gives a source term in the Maxwell equation, and it is electrically charged. 
The first term in Eq.~(\ref{eq:L_eff_WZW2}) is  proportional to the lump charge 
density while the second term is to  
the soliton charge.
In the case of a Skyrmion, 
 the second term is a total derivative (for constant $A_0$)  
and does not contribute to the total electric charge. 
Then, one finds that a contribution to the electric charge 
of (anti-)Skyrmion  
is\footnote{
From the Gell-Mann Nishijima formula, 
$I_3 + \frac{1}{2} N_{\rm B}^{\rm Sk}$  
with an isospin $I_3$  
gives an electric charge upon the quantization.
}
\begin{equation}
 Q_e^{\rm Sk} = \frac{1}{2} N_{\rm B}^{\rm Sk} = -  N_{\rm lump}  .
\end{equation}

Finally, we gather ${\cal L}_{\rm ChPT}^{\rm eff}$ given in Eq.~(\ref{eq:L_eff_ChPT2}) and the contributions from the WZW term given in Eqs.~(\ref{eq:L_eff_WZW}) and (\ref{eq:L_eff_WZW2}) to obtain the low energy effective Lagrangian on the CSL background as
\begin{eqnarray}
    {\cal L}^{\rm eff} &=& 
    - \sigma_{\rm CSL}(\kappa) - \frac{eB^z\mu_{\rm B}}{2\pi}
    - \frac{\ell}{4}F_{\alpha\beta}F^{\alpha\beta}
+ {\cal C}(\kappa)   
\left[D_\alpha \phi^\dag D^\alpha \phi + (\phi^\dag D_\alpha\phi) (\phi^\dag D^\alpha \phi)\right] \nonumber\\
&-& \left(\mu_{\rm B}+ \frac{e}{2}A_0\right)\left\{ 2 q 
+ \frac{e}{2\pi} 
\epsilon^{ab}\p_a\left(A_b{\rm tr}\left[\tau^3 P\right]\right)
\right\}\,.
\label{eq:CHPT+WZW_CSL_phi}
\end{eqnarray}
Here, we have included 
the gauge kinetic term
as the third term, 
 which is not localized on the soliton and is 
proportional to the period $\ell$.

\subsubsection{Reformulation as $O(3)$ nonlinear sigma model}
The ${\mathbb C}P^1$ model is equivalent to the 
$O(3)$ nonlinear sigma model because of ${\mathbb C}P^1\simeq S^2 \simeq O(3)/O(2)$.
Here we rewrite the above Lagrangian 
as the $O(3)$ model described by three component real scalar fields $\vec n$  defined by
\be
\vec n = \phi^\dag \vec \tau \phi\,,\quad \left[\vec n = \,{\rm tr}\left(\vec\tau P\right)\,,\quad P = \phi\phi^\dag = \frac{1}{2}\left(I_2 + \vec n \cdot \vec \tau\right)\right]\,.
\ee
This satisfies the constraint $\vec n^2 = 1$.

The electromagnetic $U(1)_{\rm EM}$ gauge transformation 
of $\vec n$ fields can be read from Eqs.~(\ref{eq:EM_transf}) and (\ref{eq:EM_transf2}) as 
\be
n_1 + i n_2 \to e^{-i\alpha(x)}(n_1+in_2)\,,\quad  
n_3 \to n_3\,.
\label{eq:EM_n}
\ee
Note that $n_1 + i n_2$ is negatively charged under $U(1)_{\rm EM}$. Indeed, $n_1 \pm i n_2$ is essentially identified with the charged pion $\pi^{\mp}$ as
\begin{eqnarray}
    \pi^\pm 
    = \frac{f_\pi}{2i}{\rm tr}\left[(\tau^1\mp i\tau^2)\Sigma\right] 
= \frac{f_\pi(u-u^{-1})}{2i}\left(n_1\mp i n_2\right)
= f_\pi \left(n_1\mp i n_2\right) \sin\chi^3_{\rm CSL}\,,
\label{eq:pi_pm}
\end{eqnarray}
where we have used Eqs.~(\ref{eq:Sigma_pion}) and (\ref{eq:Sigma}).

Similarly, we have the following expression for the neutral components
\be
\sigma = f_\pi \cos\chi^3_{\rm CSL}\,,\quad
\pi^0 = f_\pi n_3 \sin\chi^3_{\rm CSL}\,.
\label{eq:pi_0}
\ee
Correspondingly, the covariant derivatives are translated as 
\be
D_\alpha \phi = \p_\alpha \phi + i \frac{e}{2}A_\alpha \tau^3 \phi
\quad \Leftrightarrow \quad
\left\{
\begin{array}{l}
D_\alpha (n_1+i n_2) = (\partial_\alpha - i e A_\alpha)(n_1+i n_2)\\ 
D_\alpha n^3 = \p_\alpha n^3
\end{array}
\right.
\label{eq:D_n}
\ee
Then one can easily show the following expression 
\be
D_\alpha \phi^\dag D^\alpha \phi + (\phi^\dag D_\alpha\phi) (\phi^\dag D^\alpha \phi)
= \frac{1}{4}D_\alpha \vec n \cdot D^\alpha \vec n\,.
\ee
Thus the effective Lagrangian with respect to $\vec n$ field is given by
\begin{eqnarray}
    {\cal L}^{\rm eff} &=& - \sigma_{\rm CSL} 
    - \frac{eB^z\mu_{\rm B}}{2\pi}
    - \frac{\ell}{4}F_{\alpha\beta}F^{\alpha\beta} + 
    \frac{{\cal C}}{4} D_\alpha \vec n \cdot D^\alpha \vec n \nonumber\\
    &-& \left(\mu_{\rm B}+ \frac{e}{2}A_0\right)\left\{ 2 q 
+ \frac{e}{2\pi} 
\epsilon^{ab}\p_a\left(A_b n_3\right)
\right\}\,,
\label{eq:CHPT+WZW_CSL}
\end{eqnarray}
where the lump charge $q$ given by Eq.~(\ref{eq:lump_charge_q}) can be rewritten in terms of $\vec{n}$ as
\begin{eqnarray}
    q = \frac{1}{8\pi} \epsilon^{ij}\vec n\cdot (\partial_i \vec n\times\partial_j\vec n) \,.
\end{eqnarray}
The last term in 
Eq.~(\ref{eq:CHPT+WZW_CSL}), 
\begin{eqnarray}
 - \frac{e}{2}A_0 \left\{ 2 q 
+ \frac{e}{2\pi} 
\epsilon^{ab}\p_a\left(A_b n_3\right)
\right\}\,,
\label{eq:charge}
\end{eqnarray}
implies that 
an electric charge is induced 
around a lump 
as mentioned above.

It is worth making a comment on similarity between the our effective theory and magnetic systems.
If we force the $\vec n$ field to be constant and the gauge field is the uniform background $A_a = {\cal A}_a$, the Hamiltonian reduces to the following form
\begin{eqnarray}
    {\cal H}^{\rm eff} 
    \to \left(\sigma_{\rm CSL} + \frac{\ell}{2}(B^z)^2\right) + \frac{{\cal C}e^2(B^{z})^2}{16} (x^2 + y^2)(1-n_3^2)
    + \frac{eB^z\mu_{\rm B}}{2\pi} \left(1-n_3\right)\,.
    \label{eq:H_N_const}
\end{eqnarray}
The second term is the contribution from the kinetic term $(D\vec n)^2$, and the third term originates from the WZW term [the second line of Eq.~(\ref{eq:CHPT+WZW_CSL})]. The former resembles the easy-axis potential that favors $n_3 = \pm 1$. The latter is the same as the Zeeman-type potential that favors $n_3 = +1$. Thus, our effective Lagrangian is similar to a ferromagnet.
The minimization of the energy at $n_3=1$ is of course expected from a view point of CSL.
Recall $n_3 = 1$ corresponds to $\phi = (1,0)^T$ 
and 
$g={\bf 1} \in SU(2)$ in Eq.~(\ref{eq:Sigma_g}), 
implying $\Sigma_0$ in Eq.~(\ref{eq:Sigma_g}) which is the most stable CSL background.

\section{Baryons on the chiral soliton lattice}
\label{sec:phase-boundary}

In this section, 
we investigate baryons 
as Skyrmions in the CSL.
We first study (anti-) lumps in BPS approximations in Subsec.~\ref{sec:BPS} 
which is mainly a review.
In Subsec.~\ref{sec:BPS_approx}, 
we construct gauged anti-lumps as baryons
without BPS approximation.
We then discuss implications to the phase diagram
in Subsec.~\ref{sec:phase-diagram}.
In Subsec.~\ref{sec:regular}, we study  properties of regular solutions such as induced electric charge, 
baryon number and energy densities.
In Subsec.~\ref{sec:anti-baryon},
we discuss gauged lumps as anti-baryons, which are excited states.

\subsection{The $\mathbb{C}P^1$ lumps in the BPS approximation}\label{sec:BPS}

If we ignore the electromagnetic interaction in Eq.~(\ref{eq:CHPT+WZW_CSL_phi}), we have
the simple $2+1$ dimensional $\mathbb{C}P^1$ model which admits a BPS state through well-known Bogomol'nyi completion of the energy density
\begin{eqnarray}
\p_a \phi^\dag \p_a \phi + (\phi^\dag \p_a\phi) (\phi^\dag \p_a \phi)
=
\frac{\p_a u\p_a \bar u}{(1+|u|^2)^2}
= \frac{4|\bar\p u|^2}{(1+|u|^2)^2} + 2\pi q
\ge 2\pi q\,,
\end{eqnarray}
where we have introduced the inhomogeneous coordinate $u \in \mathbb{C}$ of $\mathbb{C}P^1$, defined by
\be
\phi = \frac{1}{\sqrt{1+|u|^2}}
\begin{pmatrix}
    1\\
    u
\end{pmatrix}\,,\qquad w = x^1 + i x^2\,,\quad\bar\p = \frac{\p_1+i\p_2}{2}\,.
\ee
Here $q$ is the lump charge density which is expressed in terms of $u$ as
\be
q 
= - \frac{i}{2\pi}\epsilon^{ij}\del_i\phi^{\dag}\del_j\phi
= \frac{i \p_1 u \p_2 \bar u - i \p_2 u \p_1 \bar u}{2\pi(1+|u|^2)^2}\,.
\ee

\subsubsection{BPS lumps}
\label{sec:BPS_lumps}
Let us consider $k$ ($>0$) BPS lumps with \cite{Polyakov:1975yp}
\be
u = \frac{b_{k-1}w^{k-1} + b_{k-2}w^{k-2} + \cdots + b_1 w_1 + b_0}{w^k + a_{k-1}w^{k-1} + a_{k-2}w^{k-2} + \cdots + a_1 w + a_0}\,.
\label{eq:u}
\ee
The topological charge $N_{\rm lump}$ counts exactly $k$ as 
\be
N_{\rm lump} = \int d^2x\, q = k\,.
\ee
Note that $u$ asymptotically goes to
$u \to 0$ for $|w| \to \infty$, implying
$\phi^T \to (1,0)$ and $n_3 \to 1$.

Now we calculate mass of the lumps by integrating the Hamiltonian with the background field ${\cal A}_a$ is re-included over $x^1$-$x^2$ plane:
\begin{eqnarray}
M_k^{\text{(lump)}} &=& \int d^2x\, \left[
\left(\sigma_{\rm CSL} + \frac{\ell}{2}(B^z)^2\right) + 
{\cal C} \frac{|D_au|^2}{(1+|u|^2)^2} 
    - \frac{e\mu_{\rm B}}{2\pi} \p_a\left\{
\epsilon^{ab}{\cal A}_b\left(1-n_3\right)\right\}
+ 2 \mu_{\rm B} q\right] \nonumber\\
&=& \left(\sigma_{\rm CSL} + \frac{\ell}{2}(B^z)^2\right)  S + 
2\pi {\cal C} k 
+ eB^z\mu_{\rm B} |b_{k-1}|^2 + 2\mu_{\rm B} k \nonumber\\
&+&  {\cal C} \int d^2x \frac{-ie{\cal A}_a(u\p_a\bar u - \bar u \p_a u) + e^2 {\cal A}_a^2 |u|^2}{(1+|u|^2)^2}\,,
\label{eq:ene_BPS}
\end{eqnarray}
with
\be
D_a u = (\p_a - i e{\cal A}_a)u\,
\ee
and $S$ an infinite area of the $xy$ plane.

We have 
evaluated the third term in the first line 
of Eq.~(\ref{eq:ene_BPS}) 
as
\begin{eqnarray}
- \frac{e\mu_{\rm B}}{2\pi} \oint dS_a\, \epsilon^{ab}{\cal A}_b(1-n_3)
= eB^z\mu_{\rm B} |b_{k-1}|^2\,,
\end{eqnarray}
where we have used the cylindrical coordinate $w = r e^{i\theta}$, and then we have
$dS_1 = r \cos\theta d\theta$ and
$dS_2 = r \sin\theta d\theta$.
Together with the background gauge field  given by ${\cal A}_a =\epsilon_{ab}x^bB^z/2$ as given in Eq.~(\ref{eq:A_bkg}),
we have
$dS_a\,\epsilon^{ab}{\cal A}_b = - \frac{B^z}{2}r^2$.
Furthermore, we have
$n_3 = \frac{1-|u|^2}{1+|u|^2}$,
so that
$
1-n_3 = \frac{2|u|^2}{1+|u|^2} \to \frac{2|b_{k-1}/w|^2}{1+|b_{k-1}/w|^2}
\sim \frac{2|b_{k-1}|^2}{r^2}
$ as $r \to \infty$.

The last term in Eq.~(\ref{eq:ene_BPS}) includes divergence, which is the cost of ignoring dynamics of the gauge field. For the simplest case of $k=1$, we find
the integral 
\be
{\cal C}\left[
e B^z \pi |b_0|^2\left(1+\log\frac{|b_0|^2}{|b_0|^2+\Lambda^2}\right)
+ \frac{e^2 (B^z)^2\pi |b_0|^2}{4}\left(|b_0|^2 + \Lambda^2 + 2|b_0|^2\log\frac{|b_0|^2}{|b_0|^2 + \Lambda^2}\right)
\right]\,,
\ee
with the IR cutoff $\Lambda$.
Here we temporally turn off this infinity by hand. In the next subsection, we will resolve the EOMs with including the dynamical gauge field, and show the divergence will disappear.

We thus obtain the mass difference between the CSL with and without the BPS $k$ lumps
\be
\delta M^{\rm (lump)}_k = 2\left( \pi  {\cal C}(\kappa) + \mu_{\rm B}\right) k + eB^z\mu_{\rm B} |b_{k-1}|^2\,.
\label{eq:mass_k_lump}
\ee
The last term is positive semi-definite (since the background CSL is fixed to have $n_3 = +1$, the background magnetic field should be $B^z \ge 0$), so we should set $b_{k-1} = 0$ by energy minimization. 
This implies the minimum winding configuration ($k=1$) goes to 
$b_0 \to 0$, implying 
the size modulus $|b_0|$ vanishes: 
the {\it small lump}  singularity. On the other hand,  the higher winding lumps with $k\ge2$ 
can be regular with a finite size. After all we find that creating the BPS lumps on the CSL costs positive energy, $\delta M^{\rm (BPS)}_k = 2\left( \pi  {\cal C}(\kappa) + \mu_{\rm B}\right) k \ge 0$ for $k\ge 0$. 
This is fully consistent with the fact that the corresponding baryon number is $N_{\rm B}^{\rm Sk} = - 2 N_{\rm lump} = - 2k < 0$, implying anti-baryons, as seen in Eq.~(\ref{eq:Nb_Nl}). Having negative baryon charge under a positive baryon chemical potential is energetically disfavored. 

\subsubsection{BPS anti-lumps}
Let us next consider $k$ ($>0$) BPS anti-lumps by just replacing $w$ and $\bar\p$ by $\bar w$ and $\p$, respectively. The $k$ BPS anti-lumps satisfying the boundary condition $n_3 \to + 1$ is 
\be
u = \frac{d_{k-1}\bar w^{k-1} + d_{k-2}\bar w^{k-2} + \cdots + d_1 \bar w_1 + d_0}{\bar w^k + c_{k-1}\bar w^{k-1} + c_{k-2}\bar w^{k-2} + \cdots + c_1 \bar w + c_0}\,,
\label{eq:u2}
\ee
with
\be
N_{\rm lump} = \int d^2x\, q = - k\,.
\ee
Repeating similar calculations as done above (ignoring the irrelevant divergence), we find
\begin{eqnarray}
M_k^{\text{(anti-lump)}} = \sigma_{\rm CSL} S + 2\pi {\cal C}(\kappa) k + eB^z\mu_{\rm B} |d_{k-1}|^2 - 2\mu_{\rm B} k\,,
\end{eqnarray}
and 
\begin{eqnarray}
\delta M_k^{\text{(anti-lump)}} =  2\left(\pi {\cal C}(\kappa) - \mu_{\rm B} \right) k + eB^z\mu_{\rm B} |d_{k-1}|^2
\ge 2\left(\pi {\cal C}(\kappa) - \mu_{\rm B} \right) k\,.
\label{eq:M_BPS_antilump}
\end{eqnarray}
Thus, we again need to set $d_{k-1} = 0$ for the energy minimization.

A significant difference between the BPS and anti-BPS cases is signature of the $2\mu_{\rm B}k$ in $\delta M_k$.
Due to the minus sign, emergence of the BPS anti-lumps reduces total mass of the excitations on the CSL. 
It eventually becomes negative for
\be
\mu_{\rm B} \ge \pi  {\cal C}(\kappa) = \frac{16 \pi f_\pi^2}{3m_\pi} \beta(\kappa)\,
\ee
giving 
the phase boundary 
between DWSk and CSL phases 
in the $\mu_{\rm B}$-$eB^z$ plane \cite{Eto:2023wul}. 
The elliptic modulus $\kappa$ is determined by
\begin{eqnarray}
    \mu_{\rm B}(\kappa) = \frac{16 \pi f_\pi^2}{3m_\pi} \beta(\kappa)\,,\quad
    eB^z(\kappa) 
    = 3m_\pi^2 \frac{E(\kappa)}{\kappa \beta(\kappa)}\,.
     \label{eq:PB_BPS}
\end{eqnarray}
where we have used Eq.~(\ref{eq:minimizing_cond}) for the second equation.
The phase boundary is denoted by the green curve in Fig.~\ref{fig:phase_diagram_AEN0}.

We should note that this is the result by using the BPS approximation with ignoring the dynamical gauge fields. Moreover, the $k=1$ anti-lump ($N_{\rm lump} = +1$) hits the small lump singularity. 
We will investigate these issues in more details in the subsequent subsections.

\subsection{Gauged $\mathbb{C}P^1$ anti-lumps (baryons) beyond the BPS approximation}
\label{sec:BPS_approx}

Let us go back to the Lagrangian (\ref{eq:CHPT+WZW_CSL}) and we treat the gauge field as a dynamical field. 
We first ignore the electric potential $A_0$ and focus on static and magnetic configurations.
Thus, the Lagrangian we will investigate in this subsection is given by
\begin{eqnarray}
    {\cal L}_{\rm eff}^{\mathbb{C}P^1} =  
    - \frac{\ell}{4}F_{\alpha\beta}F^{\alpha\beta} + 
    \frac{{\cal C}}{4} D_\alpha \vec n \cdot D^\alpha \vec n
    - \mu_{\rm B}\left\{ 
    2 q + \frac{e}{2\pi} \epsilon^{ab}\p_a\left(A_b (n_3-1)\right)
    \right\}\,,
\label{eq:L_eff}
\end{eqnarray}
where we have suppressed the constant $-\sigma_{\rm CSL}$ and sent 
the other constant, the second term of Eq.~(\ref{eq:CHPT+WZW_CSL}), 
into the last term.
 Note that the last term does not affect the EOMs because it is a surface topological term.
Apart form that topological term, 
this is a $U(1)$ gauged $\mathbb{C}P^1$ model. 
The Hamiltonian for a static and magnetic configuration reads
\be
{\cal H}_{\rm eff}^{\mathbb{C}P^1} = 
    \frac{\ell}{2}F_{12}^2 + \frac{{\cal C}}{4}D_a\vec n\cdot D_a \vec n 
    + \mu_{\rm B}\left\{ 
    2 q + \frac{e}{2\pi} \epsilon^{ab}\p_a\left(A_b (n_3-1)\right)
    \right\}\,.
    \label{eq:H_eff}
\ee

We shall decompose the gauge field into the background part ${\cal A}_\alpha$ and the dynamical part $a_\alpha$ as
\be
A_\alpha = {\cal A}_\alpha + a_\alpha\,,
\ee
where the background part is fixed to be
${\cal A}_a =\epsilon_{ab}x^bB^z/2$ and ${\cal A}_0 = 0$.
Then, the first two terms of Hamiltonian are decomposed as
\be
\frac{\ell}{2}F_{12}^2 + \frac{{\cal C}}{4}D_a\vec n\cdot D_a \vec n = E^{(4)} + E^{(2)} + E^{(1)} + E^{(0)}\,,
\ee
with
\begin{eqnarray}
E^{(4)} &=& \frac{\ell}{2} \int d^2x\, f_{12}^2\,,\\
E^{(2)} &=& 
\int d^2x\, \left[\ell B f_{12} + \frac{\cal C}{4}\left\{|\partial_i \Vec{n}|^2 +2e a_i\varepsilon_{ab}\partial_in_an_b + e^2 a_i^2 (n_1^2+n_2^2)\right\}\right]
 \,\\
E^{(1)} &=& \frac{{\cal C}}{4}\int d^2x\, \left[2e{\cal A}_i\epsilon_{ab}(\p_in_a)n_b + 2e^2 {\cal A}_i a_i(n_1^2+n_2^2)\right]\,,\\
E^{(0)} &=& \frac{{\cal C}}{4}\int d^2x\, e^2 {\cal A}_i^2 (n_1^2 + n_2^2)\,.
\end{eqnarray}
Here $f_{12} = \p_1 a_2 - \p_2 a_1$.
The Derrick's scaling argument 
\cite{Derrick:1964ww}
tells that for a regular solution to exist it should satisfy
\be
-E^{(4)} + E^{(1)} + 2E^{(0)} = 0\,.
\label{eq:Derrick}
\ee

Now we are ready to construct gauged lump solutions.
Let us make an Ansatz for the $k=1$ anti-lump. That for the scalar fields $\vec n$ is given by
\begin{eqnarray}
    n_1 = \frac{x}{r}\sin\Theta(r) \,,\quad
    n_2 = \frac{y}{r}\sin\Theta(r) \,,\quad
    n_3 = \cos\Theta(r)\,,
    \label{eq:Ansatz_n_antilump}
\end{eqnarray}
with $r = \sqrt{x^2+y^2}$.
We adopt the following boundary condition
\begin{eqnarray}
\Theta(0) = \pi\,,\quad \Theta(\infty) = 0\,,\quad \left[n_3(0) = -1\,,\quad n_3(\infty) = +1 \right]\,,
\end{eqnarray}
which meets the physical requirement $n_3=1$ at spatial infinity.
The lump topological charge density can be written as
\be
q = \frac{1}{8\pi} \epsilon^{ij}\vec n\cdot (\partial_i \vec n\times\partial_j\vec n) =  \frac{\Theta' \sin\Theta}{4\pi r}\,,
\ee
and therefore the lump charge reads
\be
N_{\rm lump} =
\int q\, d^2x 
= - \frac{\cos\Theta(\infty) - \cos\Theta(0)}{2}
= -1\,.
\ee

We also make the following Ansatz for the gauge field
\begin{eqnarray}
    a_0 = 0\,,\quad
    a_1 =  \frac{B^z a(r)}{2} y\,,\quad
    a_2 = - \frac{B^z a(r)}{2} x\,,
\end{eqnarray}
where $a_\mu$ is the dynamical gauge field.
Together with the background gauge field, the full gauge field is given by
\begin{eqnarray}
    A_0 = 0\,,\quad
    A_1 = \frac{B^z (1+a(r))}{2} y\,,\quad
    A_2 = - \frac{B^z (1+a(r))}{2} x\,.
    \label{eq:ansatz_gauge}
\end{eqnarray}
The mass dimension of $a(r)$ is zero.
The magnetic field is given by
\begin{eqnarray}
    F_{12} = - B^z (1+a) - \frac{B^z r a'}{2}\,,
\end{eqnarray}
where the prime stands for a derivative in terms of the physical coordinate $r$.
We impose the profile function $a(r)$ to approach $0$ as $r \to \infty$, so that the magnetic field asymptotically behaves as $F_{12} \to -B^z$.

Plugging Eqs.~(\ref{eq:Ansatz_n_antilump}) and (\ref{eq:ansatz_gauge}) into Eq.~(\ref{eq:L_eff}), we find the reduced Lagrangian for $\Theta(r)$ and $a(r)$ 
\begin{eqnarray}
    {\cal L}_{\rm eff} =  - \frac{\ell(B^z)^2}{2}\left(1+a + \frac{r a'}{2}\right)^2 
    - \frac{{\cal C}}{4} \left[ 
    \Theta'{}^2 + \frac{\left(2+ e B^z r^2 (1+a)\right)^2}{4r^2} \sin^2\Theta
    \right]\,.
\end{eqnarray}
Here we retain only the terms which contribute to the EOMs whereas the constant and surface terms are ignored.
The corresponding EOMs are given by
\begin{eqnarray}
    &&\Theta'' + \frac{\Theta'}{r} - \frac{\left(2 + eB^zr^2(1+a)\right)^2}{4r^2}\sin\Theta\cos\Theta = 0\,,
    \label{eq:eom_theta}\\
&&a'' + \frac{3a'}{r}  - \frac{e{\cal C}\left(2 + eB^z r^2 (1+a)\right)}{2\ell B^zr^2} \sin^2\Theta = 0\,.
\label{eq:eom_a}
\end{eqnarray}
We numerically solve these with the boundary condition for the $k=1$ anti-lump
\be
\Theta(0)=\pi\,,\quad 
\Theta(\infty) = 0\,,\quad 
a'(0) = 0\,,\quad 
a(\infty) = 0\,.
\label{eq:bc_antilump}
\ee
Since these are EOMs of the low energy effective theory on the CSL background with the elliptic modulus $\kappa$, both ${\cal C}$ and $\ell$ are determined by $\kappa$ as
\be
\ell(\kappa) = \frac{2 \kappa K(\kappa)}{m_\pi}\,,\quad
{\cal C}(\kappa) = \frac{16 f_\pi^2}{3m_\pi} \beta(\kappa)\,,
\label{eq:mu_B_CSL}
\ee
with $\beta(\kappa)$ given in Eq.~(\ref{eq:beta}).
We deal with $B^z$ as a free parameter whereas the critical baryon chemical potential is given by
\be
\mu_{\rm B}(\kappa;B^z) = 
\frac{E(\kappa)}{\kappa} \frac{16\pi f_\pi^2 m_\pi}{eB^z}\,.
\ee

The total mass including the surface terms of the $k=1$ anti-lump reads
\begin{eqnarray}
M^{\text{anti-lump}} = M_\gamma + M_\pi + M_{\rm WZW}\,,
\end{eqnarray}
with
\begin{eqnarray}
    M_\gamma &=& 2\pi \int_0^\infty dr\, r\left[ \frac{\ell(B^z)^2}{2}\left(1+a + \frac{r a'}{2}\right)^2 - \frac{\ell(B^z)^2}{2} \right]\,,\\
    M_\pi &=& \frac{{\cal C} \pi}{2} \int_0^\infty dr\, r 
     \left[ 
    \Theta'{}^2 + \frac{\left(2+ e B^z r^2 (1+a)\right)^2}{4r^2} \sin^2\Theta
    \right]\,,\\
    M_{\rm WZW} &=& - 2 \mu_{\rm B}
    + \frac{eB^z \mu_{\rm B}}{2} \left[ 
       r^2 (1+a) \sin^2\frac{\Theta}{2}
    \right]^\infty_0\,.
\end{eqnarray}
$M_\gamma$ is the energy of the magnetic field  in which we have subtracted a contribution of the uniform magnetic field $F_{12} = - B^z$. To evaluate $M_{\rm WZW}$, we need to figure out the asymptotic behavior of $\Theta$. The EOM (\ref{eq:eom_theta}) at large $r$ reduces to
\be
\Theta'' = \left(\frac{eB^z}{2}\right)^2 \Theta\,,
\ee
so that $\Theta$ exponentially fast decays to 0. Thus, the second term of $M_{\rm WZW}$ vanishes, and we have
\be
M_{\rm WZW} = - 2 \mu_{\rm B}(\kappa;B^z)\,.
\ee

For numerical analysis, let us rewrite the EOMs with respect to the dimensionless coordinate $\rho = \sqrt{eB^z}\, r$
\begin{eqnarray}
    &&\Theta'' + \frac{\Theta'}{\rho} - \frac{\left(2 + \rho^2(1+a)\right)^2}{4\rho^2}\sin\Theta\cos\Theta = 0\,,
    \label{eq:EOM1_dimless}\\
    && a'' + \frac{3a'}{\rho}  - 
    \gamma \frac{\left(2 + \rho^2 (1+a)\right)}{2 \rho^2} \sin^2\Theta = 0\,.
    \label{eq:EOM2_dimless}
\end{eqnarray}
Thus, the EOMs depend on the unique parameter
 $\gamma$ defined by 
\be
\gamma
\equiv \frac{2\alpha}{b} \frac{M_\pi^{\rm (BPS)}/m_\pi}{m_\pi \ell} =
\frac{\alpha}{b} \frac{64\pi f_\pi^2}{3m_\pi^2} \frac{\beta(\kappa)}{2\kappa K(\kappa)}\,,
\label{eq:tilde_gamma}
\ee
with 
\be
\alpha = \frac{e^2}{4\pi}\,,\quad
b \equiv \frac{eB^z}{m_\pi^2}\,,\quad
M_\pi^{\rm (BPS)} = 2\pi {\cal C}\,.
\ee
The dimensionless masses are given by
\begin{eqnarray}
    \tilde M_\gamma &=& \frac{4 \alpha}{m_\pi \ell b} \frac{M_\gamma}{m_\pi} = \int_0^\infty d\rho\, \rho\left[ \left(1+a + \frac{\rho a'}{2}\right)^2 - 1 \right]\,,\label{eq:Mg}\\
    \tilde M_\pi &=& \frac{4 \alpha}{m_\pi \ell b} \frac{M_\pi}{m_\pi} = \frac{\gamma}{2} \int_0^\infty d\rho\, \rho 
     \left[ 
    \Theta'{}^2 + \frac{\left(2+ \rho^2 (1+a)\right)^2}{4\rho^2} \sin^2\Theta
    \right]\,,\label{eq:Mpi}\\
    \tilde M_{\rm WZW} &=& \frac{4 \alpha}{m_\pi \ell b} \frac{M_{\rm WZW}}{m_\pi} = - \frac{E(\kappa)}{\kappa} \frac{4 \alpha}{m_\pi \ell b} \frac{32\pi f_\pi^2}{b m_\pi^2}\,.
\end{eqnarray}
Note that the mass of the BPS anti-lump is expressed as
\be
\tilde M_\pi^{\rm (BPS)} 
= \frac{4 \alpha}{m_\pi \ell b} \frac{M_\pi^{\rm (BPS)}}{m_\pi}
= 2 \gamma\,.
\ee
Furthermore, the terms in the Derrick's condition can be also expressed as follows:
\begin{eqnarray}
\tilde E^{(4)} &=& \frac{4 \alpha}{m_\pi \ell b} E^{(4)} = 
\int d\rho\, \rho \left(a + \frac{\rho a'}{2}\right)^2\,,
\label{eq:E4}\\
\tilde E^{(1)} &=& \frac{4 \alpha}{m_\pi \ell b} E^{(1)} = \frac{\gamma}{2} \int d\rho\, \rho 
\left(1 + \frac{a  \rho^2}{2}\right) \sin ^2\Theta\,,
\label{eq:E1}\\
\tilde E^{(0)} &=& \frac{4 \alpha}{m_\pi \ell b} E^{(0)} = \frac{\gamma}{2} \int d\rho\,  \frac{\rho^3}{4}\sin^2\Theta\,.
\label{eq:E0}
\end{eqnarray}
The Derrick's condition in terms of these dimensionless quantities is given by
\be
- \tilde E^{(4)} + \tilde E^{(1)} + 2 \tilde E^{(0)} = 0\,.
\ee

\begin{figure}[h]
    \begin{center}
    \includegraphics[width=13cm]{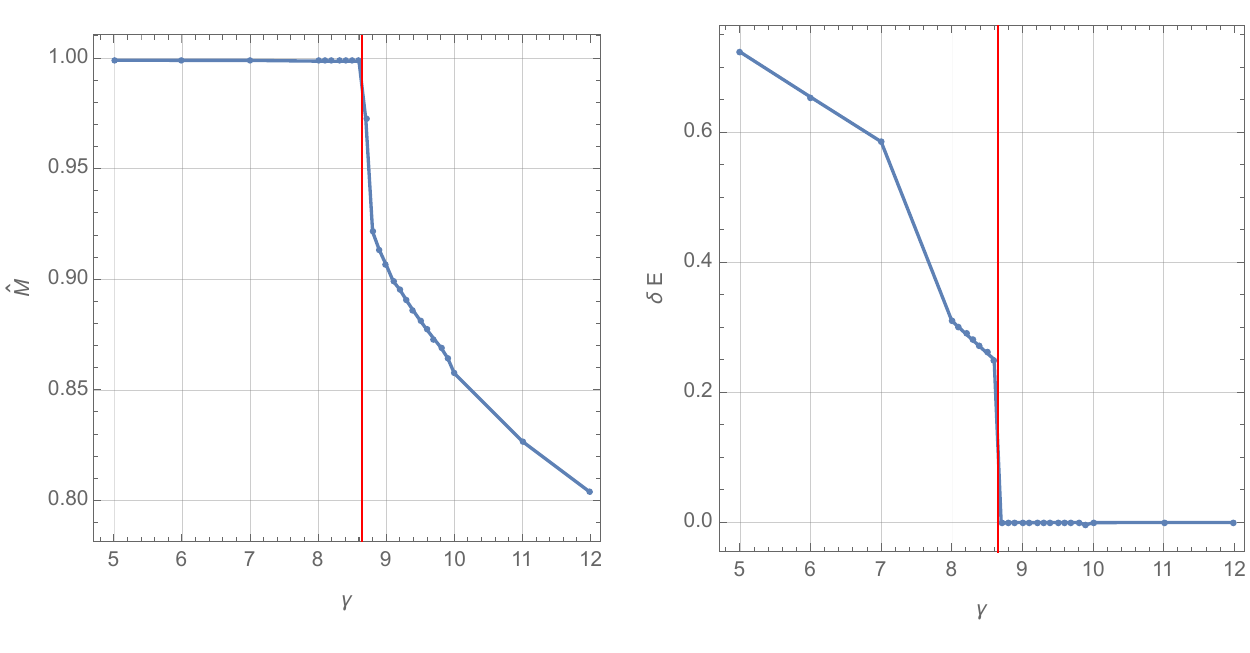}
    \caption{$\hat M$ and $\delta E$ are plotted as the functions of $ \gamma$. The red line corresponds to the critical value  $\gamma_{\rm c} \simeq 8.65$.}
    \label{fig:singular_vs_regular}
    \end{center}
\end{figure}

Now we numerically solve Eqs.~(\ref{eq:EOM1_dimless}) and (\ref{eq:EOM2_dimless}) for varying the unique parameter $ \gamma$.
We find a critical value
\be
\gamma_{\rm c} \simeq 8.65\,,
\ee
above which the numerical solutions are always regular whereas below which they collapse in a point-like configuration 
corresponding to small lumps. 
Note that reliability of our numerical solutions for $ \gamma <  \gamma_{\rm c}$ is limited because accuracy of our numerical method is limited. We will explain the small lumps in a more rigorous way below. Here, before doing that, 
let us further analyse our numerical results. 
To this end, let us define 
the following quantities:
\be
\hat M \equiv \frac{\tilde M_\gamma + \tilde M_\pi}{\tilde M_\pi^{\rm (BPS)}}\,,\quad
\delta E  \equiv \frac{- \tilde E^{(4)} + \tilde E^{(1)} + 2 \tilde E^{(0)}}{\sqrt{(\tilde E^{(4)})^2 + (\tilde E^{(1)})^2 + (\tilde E^{(0)})^2}}\,,
\label{eq:M_dE}
\ee
where $\hat M$ denotes a ratio of the mass of the gauged anti-lump to the BPS lump mass, 
and $\delta E$ measures the accuracy of the Derrick's scaling condition. 
We plot these quantities for various $ \gamma$  in Fig.~\ref{fig:singular_vs_regular}.
We can see from the right panel of 
Fig.~\ref{fig:singular_vs_regular} 
that our numerical solutions are reliable 
for $\gamma >  \gamma_{\rm c}$
where 
the Derrick's condition is fairly satisfied $\delta E \sim 0$, and we find from 
the right panel of 
Fig.~\ref{fig:singular_vs_regular} 
that $\hat M < 1$ in that region, implying that the dynamical gauge field makes the mass of $k=1$ gauged anti-lump smaller than 
that of a BPS lump. 
On the other hand, 
for $\gamma < \gamma_{\rm c}$, the numerical solution is point-like, 
where $\delta E$  significantly differs from zero and thus we cannot trust our numerical solutions.
Nevertheless, 
the behavior of the raw data of $\hat M$ is quite suggestive; it converges to $\hat M = 1$, namely the point-like solutions 
have finite energy that coincides with 
the BPS lump mass. 
This suggests that gauged anti-lumps flow to
the BPS lumps in the parameter region 
$\gamma < \gamma_{\rm c}$, 
which will be justified 
in the next subsection.

\begin{figure}[h]
    \begin{center}
    \includegraphics[width=14cm]{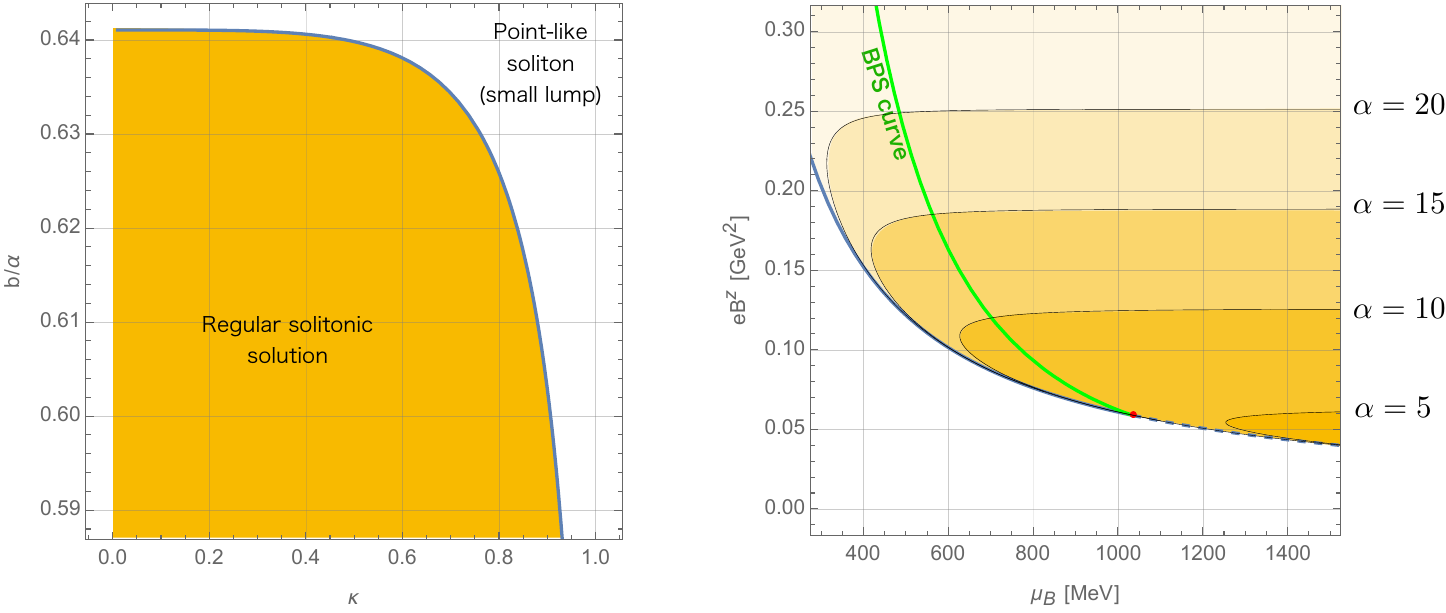}
    \caption{The phases for the regular gauged anti-lumps and point-like small anti-lumps. The left panel shows the phases in the $\kappa$-$b/\alpha$ plane. The right one shows the phases with different $\alpha$ in the $\mu_{\rm B}$-$eB^z$ plane.
    The blue solid curve at the bottom in the right panel corresponds to the CSL critical curve in Eq.~(\ref{eq:PB_SS}), the green one corresponds to the DWSk critical curve (\ref{eq:PB_BPS}) obtained in the BPS approximation, and the painted regions correspond to those that the anti-lump is a regular soliton.
    We show only relatively large $\alpha$ in $5 \le \alpha \le 20$ because the painted region quickly move down to the far right. For example, the left tip of the painted region for $\alpha = 1$ is about $\mu_{\rm B} = 6.3$ GeV and 860 GeV for $\alpha = 1/137$. We have used $m_\pi = 140$ MeV and $f_\pi = 93$ MeV for concreteness.
    }
    \label{fig:regular_singular}
    \end{center}
\end{figure}

We show in Fig.~\ref{fig:regular_singular} 
the regions where the gauged anti-lumps are regular solitonic solutions, 
and where they are point-like configurations (small lumps).
The left panel shows the relation between the elliptic modulus $\kappa$ and $b/\alpha$ given in Eq.~(\ref{eq:tilde_gamma}). The phase boundary corresponds to $\gamma = \gamma_{\rm c} = 8.56$, and we have used $m_\pi = 140$ MeV and $f_\pi = 93$ MeV for concreteness.
The right panel shows the region of regular lumps in the $\mu_{\rm B}$-$eB^z$ plane 
for various $\alpha$.


\subsection{Small lumps and the phase diagram}\label{sec:phase-diagram}

Now, we are ready to reevaluate the phase boundary between the DWSk and CSL phase, which is the main result of this paper.

In the previous subsection, we have numerically investigated Eqs.~(\ref{eq:EOM1_dimless}) and (\ref{eq:EOM2_dimless}) for the $k=1$ gauged anti-lump with the dynamical gauge field, and have found that the phase boundary between a regular soliton and a point-like soliton (small lump). Especially, we are interested in whether the anti-lump is point-like or small lump in the realistic region of $\mu_{\rm B} \sim 1$ GeV.
As shown in the right panel of Fig.~\ref{fig:regular_singular} (see also the caption), the gauged anti-lump is indeed a point-like solution for $\mu_{\rm B} \simeq 1$ GeV for the realistic value of the $U(1)_{\rm EM}$ gauge coupling: $\alpha = 1/137$.
However, as we mentioned above, since our numerical computations for the point-like solution are not so reliable, we analytically justify them here.

Our numerical solutions for the point-like configurations suggest that the dynamical gauge field is everywhere $a(r) = 0$. Thus, we will fix $a(r) = 0$, namely we will assume the dynamical gauge field is dynamically suppressed. In order to verify that the anti-lump becomes a point-like object, 
let us adapt the BPS anti-lump solution (Eq.~(\ref{eq:u2}) for $k=1$) with the size moduli $d_0$ as a variational ansatz. With respect to the $\vec n$ field, it is expressed as
\be
n_1 = \frac{2rd_0}{r^2 + d_0^2} \cos \theta\,,\quad
n_2 = \frac{2rd_0}{r^2 + d_0^2} \sin \theta\,,\quad
n_3 = \frac{r^2 - d_0^2}{r^2 + d_0^2}\,.
\ee
Then the kinetic energy density of $\vec n$ reads
\be
{\cal E}_{\vec n}
= \frac{{\cal C}}{4}D_a\vec n\cdot D_a \vec n
= \frac{{\cal C}}{4} \left\{
\frac{8 d_0^2}{\left(r^2 + d_0^2\right)^2} + 
\frac{ eB^z d_0^2 r^2 \left(4 + eB^z r^2\right)}{ \left(r^2 + d_0^2\right)^2}
\right\}
\,.
\ee
Then, the kinetic energy can be written as
\be
E_{\vec n} = F(\infty) - F(0)\,,
\ee
with
\begin{eqnarray}
F(\rho) &=& \frac{\pi{\cal C}}{4} \frac{\tilde d_0^2}{\rho^2+ \tilde d_0^2}
\left\{\rho^2 \left(\rho^2 + 2 \tilde d_0^2\right) + 2 \left(\rho^2 + \tilde d_0^2\right) \left(2- \tilde d_0^2\right) \log \frac{\rho^2 + \tilde d_0^2}{eB^z}+4 \tilde d_0^2\right\} \nonumber\\
&-& \frac{\pi{\cal C}}{4} \frac{8\tilde d_0^2}{\rho^2+ \tilde d_0^2}\,.
\end{eqnarray}
Here we have used $\tilde d_0 = \sqrt{eB^z}\, d_0$.
Introducing an IR cutoff $R$ ($\gg d_0$), we have
\be
E_{\vec n} = 2\pi{\cal C} + \frac{\pi {\cal C} eB^z d_0^2}{4}\left\{ 
eB^z R^2 
-4\log d_0^2
-2\left(2-eB^z d_0^2 \log d_0^2\right)
\right\}\,.
\ee
In the presence of the uniform magnetic field background $B^z > 0$,
this diverges in the limit $R \to \infty$ unless $d_0 = 0$.
Therefore, the size of anti-lump must be 0 for $B^z > 0$. Nevertheless, the energy remains finite even if $d_0 = 0$, and it is exactly the same as the BPS lump mass:
\be
E_{\vec n}\big|_{d_0 = 0} = 2 \pi {\cal C} = M_\pi^{\rm (BPS)}\,.
\ee
This implies that the anti-lump remains as a point-like object whose energy density is
\be
{\cal E}_{\vec n}\big|_{d_0 = 0} = 2\pi {\cal C} \delta^{(2)}(x)\,.
\ee
Including the WZW term, the total energy of the $k=1$ anti-lump 
is given by
\be
E_{k=1} =  2\pi {\cal C} - 2 \mu_{\rm B}\,.
\ee
Hence, the phase boundary between CSL with/without baryons coincides with that we found in the BPS approximation
\be
\mu_{\rm B} \ge \frac{16\pi f_\pi^2}{3m_\pi}\beta(\kappa)\,,\qquad
\text{for}\quad\alpha = \frac{1}{137}\,.
\ee

The phase boundary is given by the condition that the lower bound is saturated. The value of magnetic field can be determined from Eq.~(\ref{eq:minimizing_cond}). 
We thus find the phase boundary parameterized by $\kappa$ as
    \be
    (\mu_{\rm B},eB^z) = 
    \left(
    \frac{16\pi  f_\pi^2}{3m_\pi}\beta(\kappa),
    3m_\pi^2 \frac{E(\kappa)}{\kappa \beta(\kappa)}
    \right)\,.
    \label{eq:PB_new}
    \ee
    We show the phase diagram in Fig.~\ref{fig:phase_diagram_AEN0}.

Let us make comments on comparisons with the previous studies.
One should not think that this is just a repetition of the previous work \cite{Eto:2023lyo} which we have reviewed in Sec.~\ref{sec:BPS_approx}, though the final conclusion is unchanged. In the previous work \cite{Eto:2023lyo} and Sec.~\ref{sec:BPS_approx}, we have entirely ignored both the background and dynamical gauge fields for constructing the lump solutions. Then we made use of the BPS approximation and reached at the conclusion that the anti-lump is a point-like solution due to the WZW term as explained in Eq.~(\ref{eq:M_BPS_antilump}). Especially, that result was independent of $e$ and $B^z$ separately but dependent of $eB^z$, because they only appear in the pair of $eB^z$ when we omit the kinetic term of the gauge field $(F_{\mu\nu})^2/4$ in the Lagrangian. In contrast, in this work we have refined the previous studies in Ref.~\cite{Eto:2023lyo} by including the dynamical gauge field $a$. We confirmed that the anti-lump is a point-like object but it is not due to the WZW term but the dynamics of the $\vec n$ and the gauge field. Particularly, this conclusion depends on $\alpha$ in contrast to the previous analysis. If we take a large value for $\alpha$ than $1/137$, the anti-lump could be a solitonic solution with a finite size as we will briefly discuss in the next subsection.

\subsection{Properties of regular gauged anti-lump solutions} \label{sec:regular}

\subsubsection{Baryon number and energy densities}
\begin{figure}[h]
    \begin{center}
    \includegraphics[width=12cm]{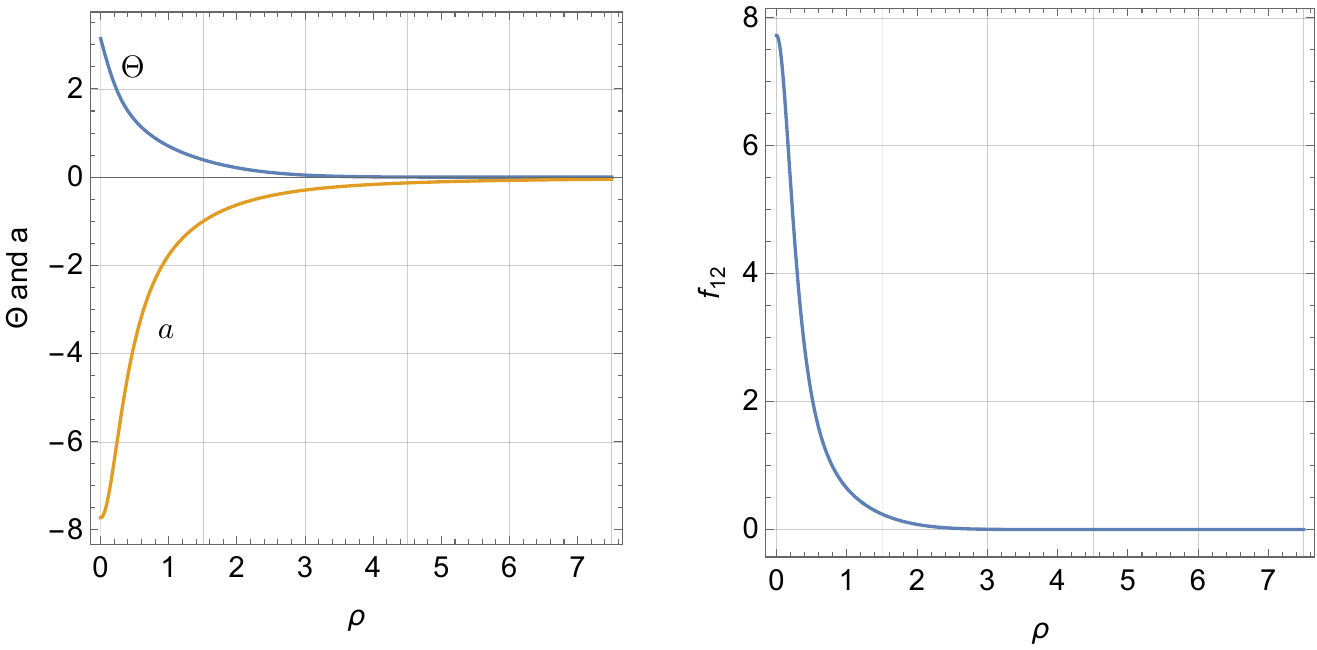}
    \caption{The profiles of the numerical solution of $k=1$ anti-lump for $\gamma = 10$.}
    \label{fig:sol_alpha_10}
    \end{center}
\end{figure}

We investigate the regular 
lump solutions in this subsection.
For that purpose, we take $\gamma$  larger than $\gamma_{\rm c}$. We choose $\gamma = 10$ as a reference value. The corresponding profiles of the numerical solution are given in Fig.~\ref{fig:sol_alpha_10}. 

Let us also show 3 dimensional visualization of the domain-wall Skyrmions.
To this end, let us first define the energy density of the CSL  
by substituting the CSL solution to 
Eq.~(\ref{eq:energy_functional}), 
which can be decomposed into contributions from the chiral Lagrangian and the WZW term:
\begin{eqnarray}
{\cal M}_{\rm CSL} = {\cal M}_{\rm CSL}^{(\Sigma)} + {\cal M}_{\rm CSL}^{\rm (WZW)}\,.
\end{eqnarray}
Here, the first term is a contribution from the chiral Lagrangian 
\begin{eqnarray}
\frac{{\cal M}^{(\Sigma)}_{\rm CSL}}{f_\pi^2 m_\pi^2}
=  \frac{1}{2}\left(\frac{d\chi_3(\tilde z)}{d\tilde z}\right)^2 - \left( \cos\chi_3(\tilde z) - 1\right)\,,
\end{eqnarray}
where we have introduced $\tilde z = m_\pi z$ and $\chi_3(\tilde z) = 2\,{\rm am}\left(\frac{\tilde z}{\kappa},\kappa\right) + \pi$. 
The second term is a contribution from the WZW term: 
\begin{eqnarray}
{\cal M}_{\rm CSL}^{\rm (WZW)} 
= 
\mu_{\rm B} j_{{\rm B;CSL}}^0 
\end{eqnarray}
where 
$j_{{\rm B;CSL}}^0$ is
the baryon charge density 
in Eq.~(\ref{eq:JB0}), evaluated 
in the CSL background as
\begin{eqnarray}
j_{{\rm B;CSL}}^0 = 
- {\cal B}_{\rm CSL}
+ \tilde{\cal B}_{\rm CSL} 
= -\frac{eB^z m_\pi}{4 \pi ^2} \frac{d\chi_3(\tilde z)}{d\tilde z}\,
\end{eqnarray}
with ${\cal B}_{\rm CSL} = 0$.
Then, we have
\be
\frac{{\cal M}_{\rm CSL}^{\rm (WZW)}}{f_\pi^2 m_\pi^2} = - \frac{4E(\kappa)}{\pi\kappa} \frac{d\chi_3(\tilde z)}{d\tilde z}\,.
\ee
Note ${\cal M}_{\rm CSL}$ is dependent of $\tilde z$ only.

The total energy density ${\cal M}_{\rm DWSk}$ 
for the domain-wall Skyrmion 
can be decomposed into a contribution from 
the CSL background ${\cal M}_{\rm CSL}$ and one from the Skyrmion ${\cal M}_{\rm Sk}$: 
\begin{eqnarray}
{\cal M}_{\rm DWSk} = {\cal M}_{\rm CSL} + {\cal M}_{\rm Sk}\, ,
\end{eqnarray} 
which actually defines 
${\cal M}_{\rm Sk}$. 
Then, the Skyrmion energy can be further decomposed into  
contributions from the 
chiral Lagrangian and WZW term: 
\be
{\cal M}_{\rm Sk} = {\cal M}_{\rm Sk}^{(\Sigma)}+ {\cal M}_{\rm Sk}^{\rm (WZW)}\,.\quad 
\ee
Here the first term is a contribution from the chiral Lagrangian 
\begin{eqnarray}
\frac{{\cal M}^{(\Sigma)}_{\rm Sk}}{f_\pi^2m_\pi^2}
&=& \frac{(eB^z)^2}{8\pi \alpha f_\pi^2m_\pi^2}\left[\frac{1}{4}
\left\{\left(2 + 2a(\rho) + \rho\frac{da(\rho)}{d\rho}\right)^2 -4\right\}\right.\nonumber \\
&+&  
\left.
\frac{3\gamma}{8} \frac{\kappa K(\kappa)}{\beta(\kappa)}
\left\{
\left(\frac{d\Theta(\rho)}{d\rho}\right)^2 + \frac{
\left(2 + \rho^2(1+ a(\rho)\right)^2
\sin^2\Theta(\rho)}{4\rho^2}
\right\}
\sin^2\chi_3(\tilde z)\right]\,,
\end{eqnarray}
where we have used
$\frac{f_\pi^2 e}{B^z} = 
\frac{3\gamma}{8} \frac{\kappa K(\kappa)}{\beta(\kappa)}
$.
The second term is 
a contribution from the WZW term 
\be
{\cal M}_{\rm Sk}^{\rm (WZW)} = \mu_{\rm B}j_{{\rm B; Sk}}^0\,, 
\ee
where $j_{{\rm B; Sk}}^0$ is
the baryon charge density, 
which can be decomposed as
\be
j_{{\rm B; Sk}}^0 = - {\cal B}_{\rm Sk} + \tilde{\cal B}_{\rm Sk}\,,
\ee
with 
\begin{eqnarray}
{\cal B}_{\rm Sk} 
&=&  - \frac{eB^z m_\pi}{4 \pi^2} \frac{2}{\rho}\frac{d\Theta(\rho)}{d\rho}  \sin \Theta(\rho) \frac{d\chi_3(\tilde z)}{d\tilde z}  \sin^2\chi_3(\tilde z)\,, \\
\tilde {\cal B}_{\rm Sk}
&=& \frac{eB^zm_\pi}{4 \pi^2} \frac{1}{2}
\frac{d\chi_3(\tilde z)}{d\tilde z}\bigg[
2 \rho (1 + a(\rho)) \frac{d\Theta(\rho)}{d\rho} \sin \Theta (\rho) \sin ^2\chi_3(\tilde z) \nonumber\\
&& -\left(2 + 2a(\rho) + \rho \frac{da(\rho)}{d\rho}\right) \cos\Theta (\rho) + 2
\bigg]\,.
\end{eqnarray}
Then we have
\begin{eqnarray}
\frac{{\cal M}_{\rm Sk}^{\rm (WZW)}}{f_\pi^2 m_\pi^2} &=& 
\frac{4E(\kappa)}{\pi\kappa}
\left[
\frac{2}{\rho}\frac{d\Theta(\rho)}{d\rho}  \sin \Theta(\rho) \frac{d\chi_3(\tilde z)}{d\tilde z}  \sin^2\chi_3(\tilde z)
\right. \nonumber\\
&+& \frac{1}{2}
\frac{d\chi_3(\tilde z)}{d\tilde z}\bigg\{
2 \rho (1 + a(\rho)) \frac{d\Theta(\rho)}{d\rho} \sin \Theta (\rho) \sin ^2\chi_3(\tilde z) \nonumber\\
&& \left.
-\left(2 + 2a(\rho) + \rho \frac{da(\rho)}{d\rho}\right) \cos\Theta (\rho) + 2
\bigg\}
\right].
\end{eqnarray}

\begin{figure}[h]
    \begin{center}
    \includegraphics[width=14cm]{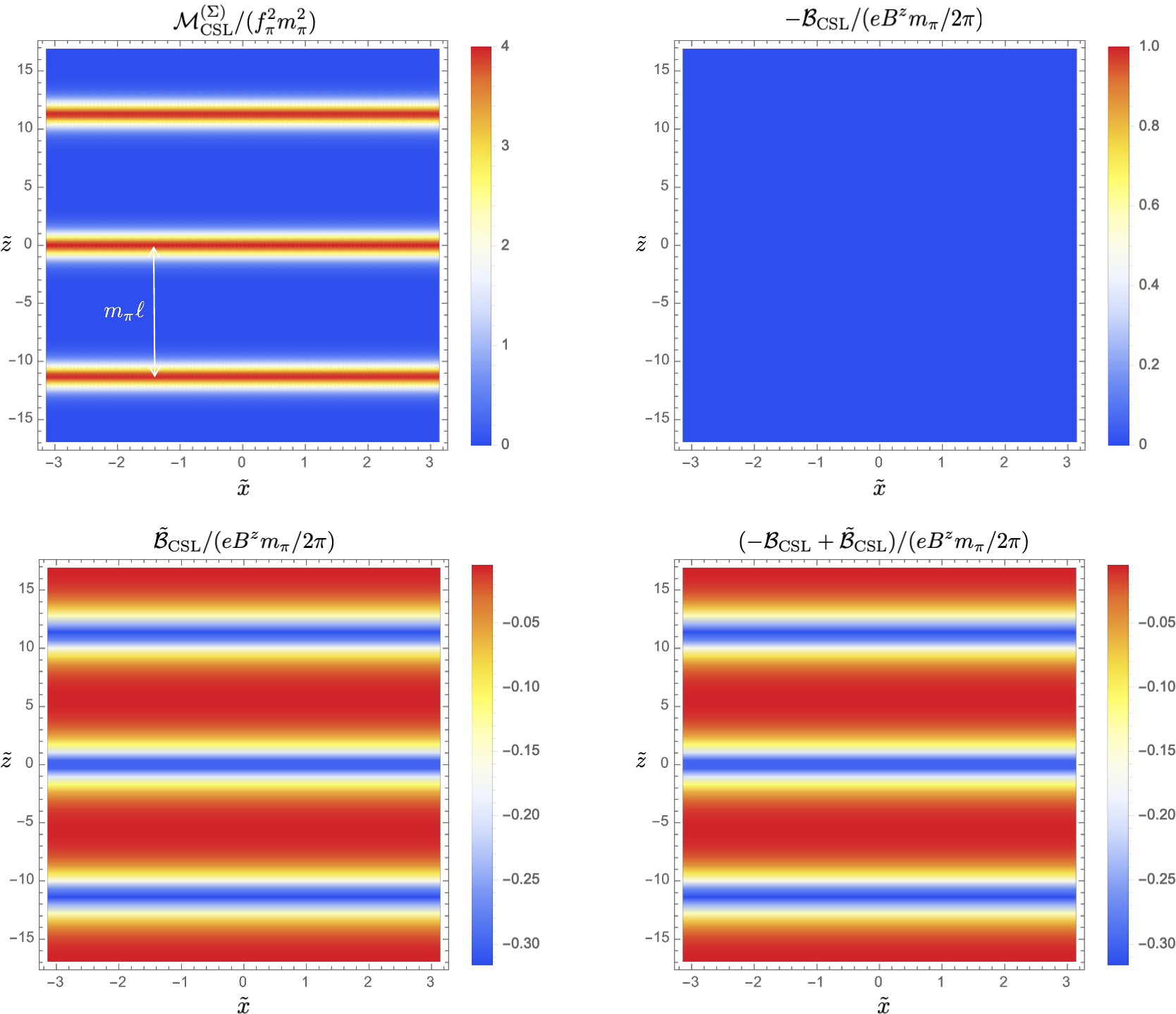}
    \caption{The cross sections at $y=0$ of dimensionless energy densities of the CSL background with $\kappa = 0.9999$ ($m_\pi\ell = 11.2892$). The kinetic energy ${\cal M}^{(\Sigma)}_{\rm CSL}$ positively contribute to the total energy density whereas the contribution of the baryon charge density $j_{\rm B;CSL}^0 = - {\cal B}_{\rm CSL} + \tilde {\cal B}_{\rm CSL}$ from the WZW term is negative. The dimensionless coordinates are defined by $\tilde x = \sqrt{eB^z}x$ and $\tilde z = m_\pi z$.}
    \label{fig:energy_densities0}
    \end{center}
\end{figure}
Now we are ready to show  configurations of a domain-wall Skyrmion for $\gamma = 10$.
We first plot in 
Fig.~\ref{fig:energy_densities0} 
a cross section at $y=0$ 
of contributions of the CSL to 
the kinetic energy ${\cal M}^{(\Sigma)}_{\rm CSL}$ 
and the baryon number densities 
$- {\cal B}_{\rm CSL}(=0)$, 
$\tilde {\cal B}_{\rm CSL}$ 
and the total baryon number density
$j_{\rm B;CSL}^0 = - {\cal B}_{\rm CSL} + \tilde {\cal B}_{\rm CSL}$. 
One can observe that 
${\cal M}^{(\Sigma)}_{\rm CSL}$
positively contributes to the total energy density whereas the contribution of the baryon charge density $j_{\rm B;CSL}^0 = - {\cal B}_{\rm CSL} + \tilde {\cal B}_{\rm CSL}$ from the WZW term 
to the total energy is negative.

\begin{figure}[h]
    \begin{center}
    \includegraphics[width=14cm]{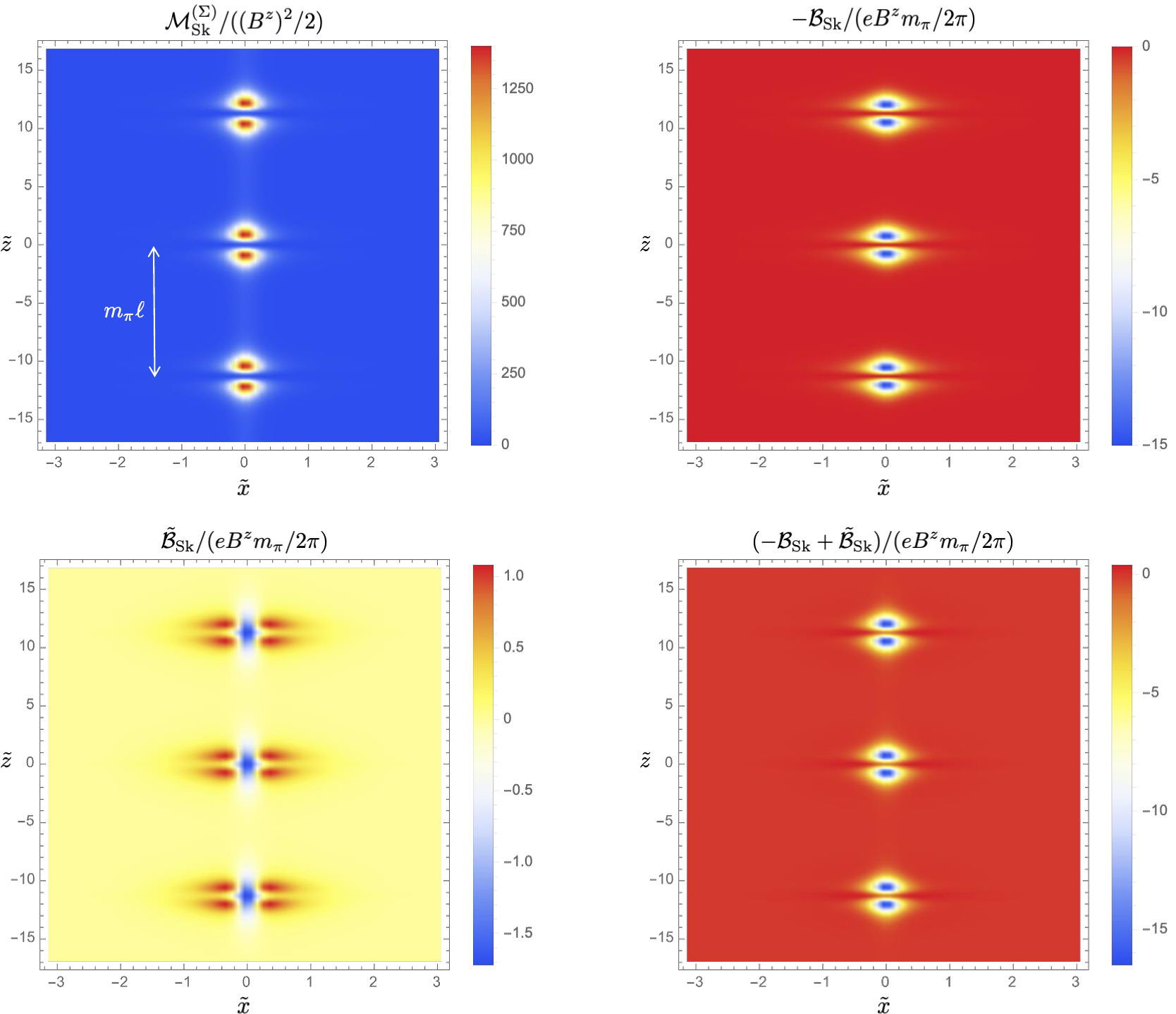}
    \caption{The cross sections at $y=0$ of dimensionless energy densities of $k=1$ anti-lump for $\gamma = 10$. We take $\kappa = 0.9999$ ($m_\pi\ell = 11.2892$). The kinetic energy ${\cal M}^{(\Sigma)}$ positively contributes to the total energy density whereas the contribution of the baryon charge density $j_{\rm B}^0 = - {\cal B}_1 + \tilde {\cal B}_1$ from the WZW term is negative (The contribution of $\tilde B_1$ vanishes because of $\int d^2x\, \tilde B_1 =0$. The dimensionless coordinates are defined by $\tilde x = \sqrt{eB^z}x$ and $\tilde z = m_\pi z$.}
    \label{fig:energy_densities1}
    \end{center}
\end{figure}
Next we plot 
in Fig.~\ref{fig:energy_densities1} 
contributions of a Skyrmion to
the kinetic energy ${\cal M}_{\rm Sk}^{(\Sigma)}$, 
the baryon number densities
$-{\cal B}_{\rm Sk}$, 
$\tilde {\cal B}_{\rm Sk}$ 
and the total baryon number density
$j_{\rm B;Sky}^0 = -{\cal B}_{\rm Sk} + \tilde {\cal B}_{\rm Sk}$. 
One can observe that 
${\cal M}^{(\Sigma)}_{\rm Sk}$
positively contributes to the total energy density whereas 
the baryon number density $j_{\rm B;CSL}^0 = - {\cal B}_{\rm Sk} + \tilde {\cal B}_{\rm Sk}$ from the WZW term 
contribute negatively to the total energy.

\begin{figure}[h]
    \begin{center}
    \includegraphics[width=15cm]{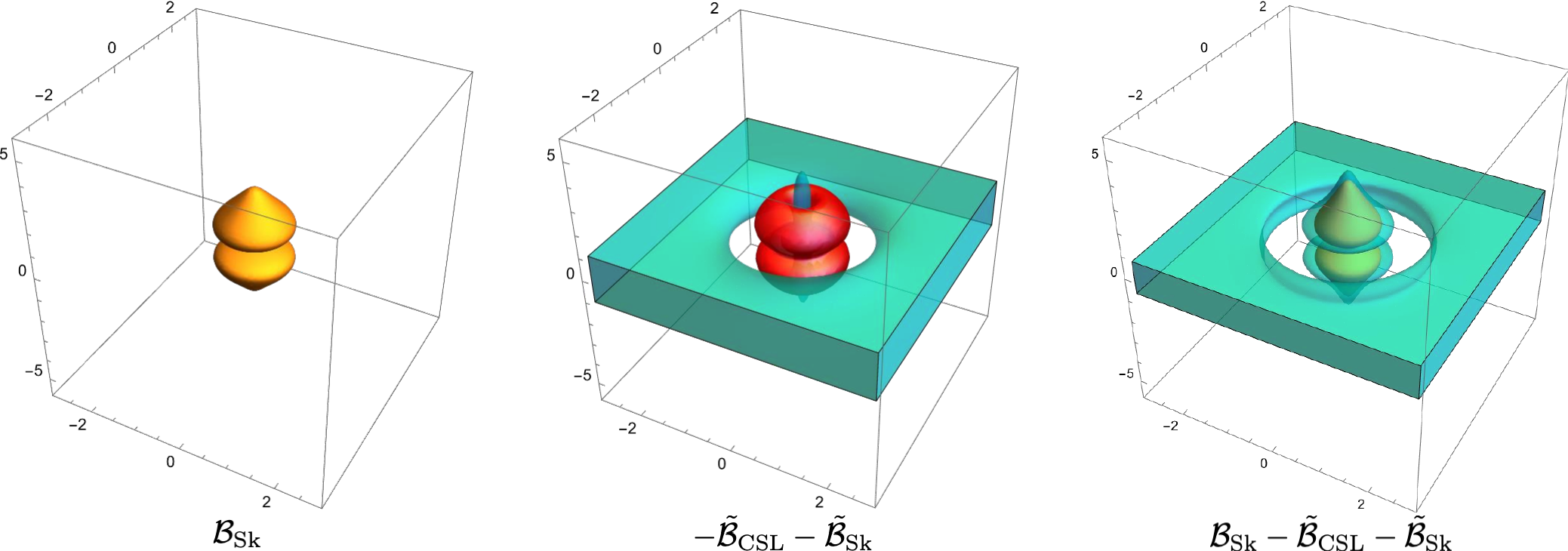}
    \caption{Isosurfaces of the baryon densities of the solitonic domain-wall Skyrmion for $\gamma = 10$ and $\kappa = 0.9999$. The left panel shows ${\cal B}_{\rm Sk}/(eB^zm_\pi/2\pi) = 0.5$. The green and red surfaces in the middle panel correspond to $-(\tilde{\cal B}_{\rm lump}+\tilde{\cal B}_{\rm Sk})/(eB^zm_\pi/2\pi) = 0.2$ and $-0.1$, respectively. The right panel shows the sum $({\cal B}_{\rm Sk}-\tilde{\cal B}_{\rm lump}-\tilde{\cal B}_{\rm Sk})/(eB^zm_\pi/2\pi) = 0.5$ (inner object) and 0.25 (green outer surface). The horizontal axes are $\tilde x$ and $\tilde y$ whereas the vertical one is $\tilde z$.}
    \label{fig:3d_fig}
    \end{center}
\end{figure}

The superposition of the contributions 
from the CSL and Skyrmion gives 
the total baryon number and energy densities.
Here, instead of do that in the 2D cross section,
we show 3D plots in Fig.~\ref{fig:3d_fig}.

\subsubsection{Phase diagram for $\alpha = 10$ }
\begin{figure}[h]
    \begin{center}
    \includegraphics[width=12cm]{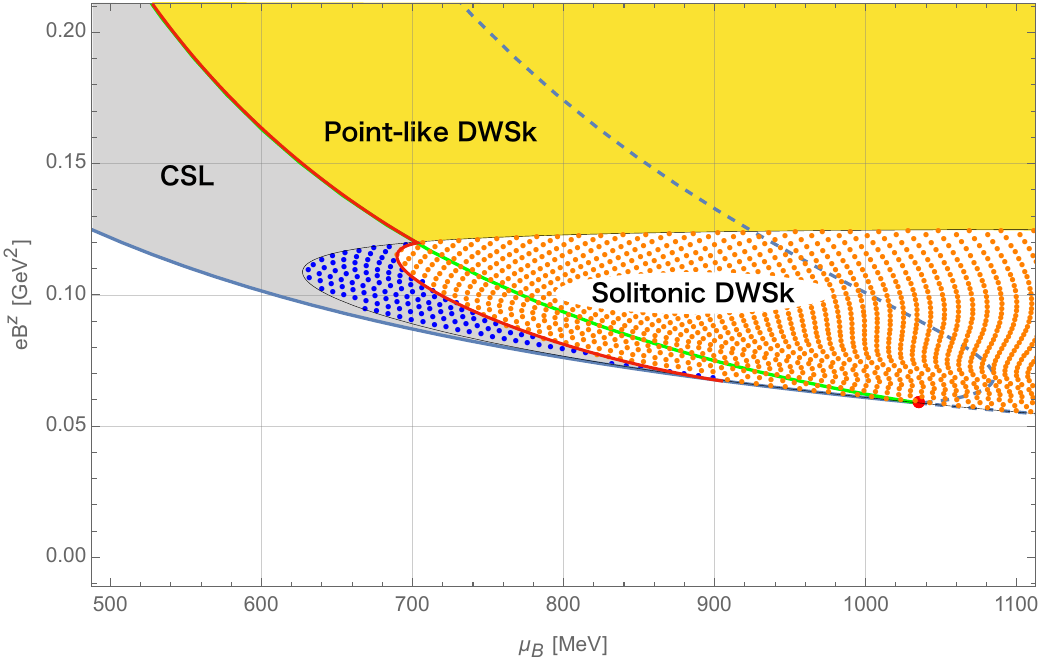}
    \caption{
    The phase diagram for $\alpha =10$. The black solid curve denotes the boundary between regular (solitonic) and point-like  (small) gauged anti-lumps. 
    The region of the solitonic DWSks is further divided into those negative energy (denoted by the orange dots) and those with positive energy (denoted by the blue dots). 
    Thus, the DWSk phase is expanded from the BPS curve (the gree solid curve) by the region with the orange dots, and the phase boundary between the DWSk and CSL phases is the red curve. The parameters are
    $\alpha = 10$, $m_\pi = 140$ MeV, $f_\pi = 93$ MeV}
    \label{fig:improved}
    \end{center}
\end{figure}

Let us show the phase diagram for $\alpha = 10$ 
in Fig.~\ref{fig:improved}, though $\alpha =10$ is not realistic value.
The phase boundary beween 
the DWSk and CSL for $\alpha = 10$ 
is now 
denoted by the red curve, 
where one can observe that 
the DWSk phase is expanded from the BPS curve (denoted by the green  solid curve). 
The black solid curve separates 
the boundary regular (solitonic) and point-like  (small) gauged anti-lumps. 
The region of the solitonic domain-wall Skyrmions is further divided into those with negative energy (denoted by the orange dots) and those with positive energy (denoted by the blue dots). 
In the former, gauge field dynamics lower the energy of gauged anti-lumps than the BPS lump mass. Therefore, the DWSk phase is expanded from the BPS curve. 
    
From this case of the strong gauge coupling, one can think it to be nontrivial that the phase diagram for the realistic gauge coupling $\alpha=1/137$ is unchanged from the BPS approximation.

\clearpage

\subsection{Gauged lumps (anti-baryons) as excited states}\label{sec:anti-baryon}

Let us next investigate $k=1$ lump with $N_{\rm lump} = 1$ which corresponds to $N_{\rm B}^{\rm Sk} = -2$ anti-baryons. Similarly to the BPS lump discussed in Sec.~\ref{sec:BPS_lumps}, it is always an excited state because contribution of the WZW term to the energy density in Eq.~(\ref{eq:H_eff}) is positive.  The Ansatz for $\Theta$ is given by
\begin{eqnarray}
    n_1 = \frac{x}{r}\sin\Theta(r) \,,\quad
    n_2 = -\frac{y}{r}\sin\Theta(r) \,,\quad
    n_3 = \cos\Theta(r)\,.
    \label{eq:ansatz_lump}
\end{eqnarray}
The boundary condition for the lump in the $n_3 = 1$ vacuum is given by
\begin{eqnarray}
\Theta(0) = \pi\,,\quad \Theta(\infty) = 0\,,\quad \left[n_3(0) = -1\,,\quad n_3(\infty) = +1 \right]\,.
\end{eqnarray}
The Ansatz for the dynamical gauge field $a$ is the same as Eq.~(\ref{eq:ansatz_gauge}).
The topological charge density is given by
\be
q  
= -  \frac{\Theta' \sin\Theta}{4\pi r}\,,
\ee
and the lump charge indeed reads
\be
N_{\rm lump} = \int q\, d^2x 
= \frac{\cos\Theta(\infty) - \cos\Theta(0)}{2}
= 1\,.
\ee

\begin{figure}[h]
\begin{center}
\includegraphics[width=12cm]{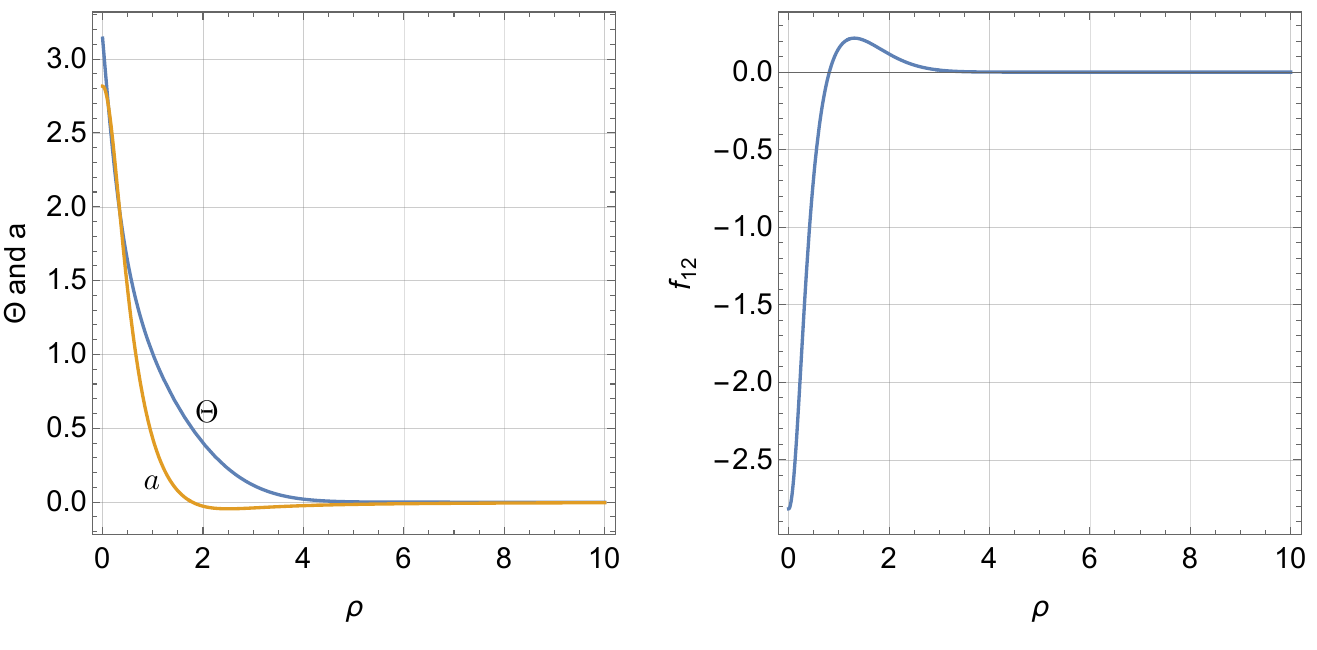}
\caption{The profiles of the numerical solution of $k=1$ lump for $\gamma = 5$.}
\label{fig:lump_gamma10}
\end{center}
\end{figure}

Note that the Ansatz in Eq.~(\ref{eq:ansatz_lump}) for the lump can be obtained from the Ansatz in Eq.~(\ref{eq:Ansatz_n_antilump}) for the anti-lump by $n_1 + i n_2 \to n_1 - i n_2$, and this is nothing but the charge conjugation $e \to - e$.
Therefore, the effective Lagrangian with the lump Ansatz reads
\begin{eqnarray}
    {\cal L}_{\rm eff} =  - \frac{\ell(B^z)^2}{2}\left(1+a + \frac{r a'}{2}\right)^2 
    - \frac{{\cal C}}{4} \left[ 
    \Theta'{}^2 + \frac{\left(2 - e B^z r^2 (1+a)\right)^2}{4r^2} \sin^2\Theta
    \right]\,.
\end{eqnarray}
Here we retain only the terms which contribute to the EOMs whereas the constant and surface terms are ignored as before.
The corresponding EOMs with respect to the dimensionless variables are given by
\begin{eqnarray}
    &&\Theta'' + \frac{\Theta'}{\rho} - \frac{\left(2 - \rho^2(1+a)\right)^2}{4\rho^2}\sin\Theta\cos\Theta = 0\,,
    \label{eq:EOM1_dimless2}\\
    && a'' + \frac{3a'}{\rho} + 
    \gamma \frac{\left(2 - \rho^2 (1+a)\right)}{2 \rho^2} \sin^2\Theta = 0\,.
    \label{eq:EOM2_dimless2}
\end{eqnarray}
As the anti-lump case, these include the unique parameter $\gamma$. We numerically solve the EOMs for varying $\gamma$. In contrast to the anti-lump solution, we find a regular solitonic lump for any $\gamma$.
Fig.~\ref{fig:lump_gamma10} shows a typical lump solution with $\gamma = 5$.
The magnetic flux confined at the center of the lump is negative, which is opposite to that of the anti-lump.

The energy of the lump is a sum of $\tilde M_\gamma$ given in Eq.~(\ref{eq:Mg}) and $\tilde M_\pi$ given by
\begin{eqnarray}
    \tilde M_\pi = \frac{4 \alpha}{m_\pi \ell b} \frac{M_\pi}{m_\pi} = \frac{\gamma}{2} \int_0^\infty d\rho\, \rho 
     \left[ 
    \Theta'{}^2 + \frac{\left(2- \rho^2 (1+a)\right)^2}{4\rho^2} \sin^2\Theta
    \right]\,.\label{eq:Mpi_lump}
\end{eqnarray}
This is slightly different from Eq.~(\ref{eq:Mpi}).
Furthermore, $\tilde E^{(4)}$ and $\tilde E^{(0)}$ in the Derrick's condition are same as those given in Eqs.~(\ref{eq:E4}) and (\ref{eq:E0}) whereas
$\tilde E^{(1)}$ differs from Eq.~(\ref{eq:E1}) and it is given by
\begin{eqnarray}
\tilde E^{(1)} = \frac{4 \alpha}{m_\pi \ell b} E^{(1)} = \frac{\gamma}{2} \int d\rho\, \rho 
\left(-1 + \frac{a  \rho^2}{2}\right) \sin ^2\Theta\,.
\end{eqnarray}

\begin{figure}[h]
    \begin{center}
    \includegraphics[width=13cm]{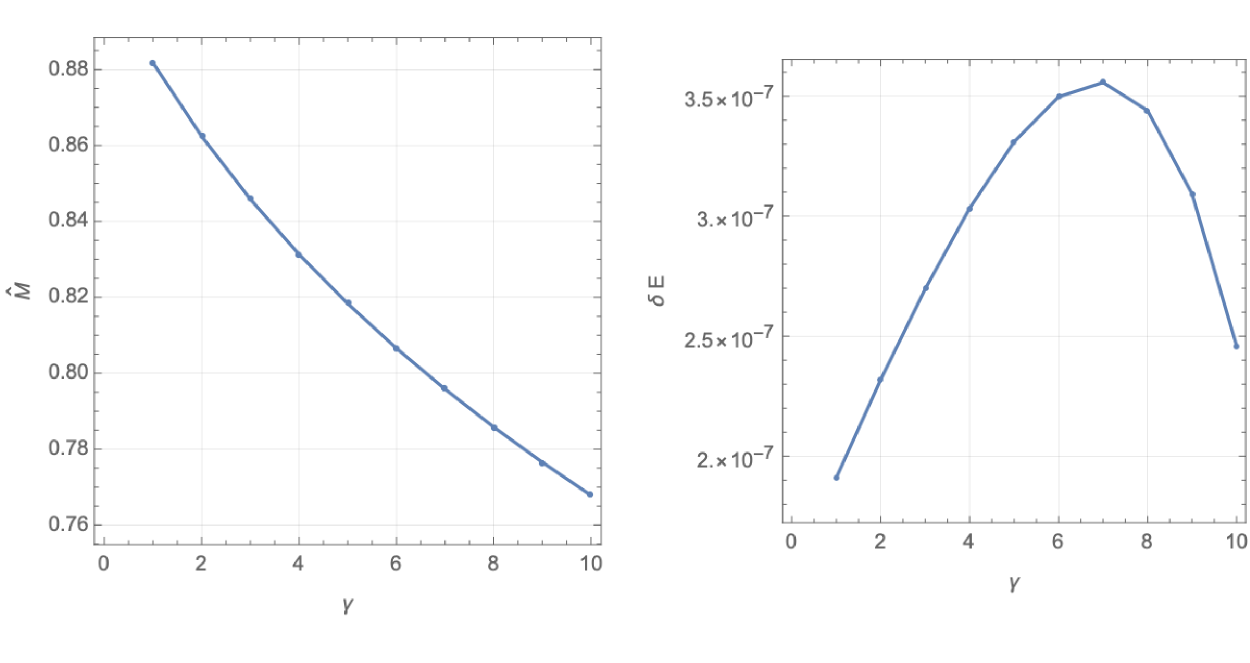}
    \caption{$\hat M$ and $\delta E$ are plotted as the functions of $ \gamma$ for a lump (anti-baryon in the bulk). }
    \label{fig:regular_lump}
    \end{center}
\end{figure}
We show $\hat M$ and $\delta E$ given in Eq.~(\ref{eq:M_dE}) for  the gauged lump in Fig.~\ref{fig:regular_lump}. 
One can see from the left panel that the lump energy is less than the BPS lump energy for any $\gamma$, and 
can check from the right panel 
that the Derrick's theorem is well satisfied in the whole region of  $\gamma$. 
This is in contrast to the case of the gauged anti-lumps (baryons) 
in Fig.~\ref{fig:singular_vs_regular}
in which the Derrick's theorem is satisfied only above the critical value $\gamma$.

\section{Summary and Discussion}
\label{sec:summary}

We have studied the phase diagram of QCD at finite baryon density with strong magnetic fields. 
We have found that 
the phase boundary between the CSL and DWSk phases previously obtained in the BPS approximation is unchanged 
beyond the BPS approximation 
taking into account the dynamics of the gauge field.
We also have found that the domain-wall Skyrmions are electrically charged with the charge one as a result of the chiral anomaly. 
For realistic value of the gauge coupling 
the anti-lumps in the domain-wall effective theory 
(baryons in the bulk) 
becomes small lumps of zero size  
but are still physical with finite energy, implying that the phase boundary is unchanged from the BPS curve. On the other hand,
 at strong gauge coupling their solutions become regular and their energy is smaller than BPS lump mass,  implying that the phase boundary changes from the BPS curve as the DWSk phase expands as in Fig.~\ref{fig:improved}. 
 While gauged anti-lumps can have negative energy due to the WZW term in the DWSk phase, 
gauged lumps (anti-baryons in the bulk) are regular and have positive energy in any parameter regions.

Let us make comments on  future directions. 
In this paper, we mainly concentrated on 
single gauged (anti-)lumps 
for the purpose of determining the phase boundary between the DWSk and CSL phases. 
In order to see a structure of DWSk phase, 
one has to discuss multiple gauged anti-lumps and investigate interaction among them.
When there are more than one anti-lumps, 
the limit of zero size modulus allows 
 finite size anti-lumps. 
In order to have a Skyrmion crystal, anti-lumps must feel repulsion.
From our previous work \cite{Amari:2024adu}
on gauged lumps in the background magnetic field,
(anti-)lumps seem to feel attraction 
(when the WZW term is neglected).
However, as shown in this paper, anti-lumps are electrically charged due to the WZW term. Thus, we can expect  a Skyrmion crystal on the soliton.

In this paper, we have used the moduli approximation, 
that does not include a back reaction from the Skyrmions 
to the solitons. 
It is an important future problem 
to construct full three dimensional solutions of 
domain-wall Skyrmions, 
including a Skyrmion lattice structure mentioned above.

When the isospin chemical potential is introduced,
the charged pions are condensed in the ground state.
In such a case, 
there appears a vortex-Skyrmion phase
\cite{Qiu:2024zpg}, 
where a vortex-Skyrmion 
is a Skyrmion hosted by a vortex 
rather than a domain wall 
\cite{Gudnason:2014hsa,Gudnason:2014jga,Gudnason:2016yix}.  
It is an open question how 
the vortex-Skyrmion phase 
and domain-wall Skyrmion phase 
are connected in a phase diagram spaned by 
the isospin chemical potential.

One of future directions is a generalization to $N_{\rm F}$ flavors.
In such a case, the effective theory on the soliton is the ${\mathbb C}P^{N_{\rm F}-1}$ model. Skyrmions in the ${\mathbb C}P^{2}$ 
 model ($N_{\rm F}=3$) with 
Dzyaloshinskii-Moriya interaction 
were studied recently \cite{Akagi:2021dpk,Amari:2022boe}.

\begin{acknowledgments}
This work is supported in part by 
 JSPS KAKENHI [Grants  No.~JP23KJ1881 (YA), No. JP22H01221 (ME and MN)] and the WPI program ``Sustainability with Knotted Chiral Meta Matter (SKCM$^2$)'' at Hiroshima University (ME and MN).
 
\end{acknowledgments}

\bibliographystyle{jhep}
\bibliography{reference}

\end{document}